\documentclass[fleqn,usenatbib,useAMS]{mnras}

\usepackage{graphicx}	
\usepackage{amsmath}	
\usepackage{amssymb}	
\usepackage{bm}		
\usepackage{pdflscape}	
\usepackage{hyperref}
\usepackage{subcaption} 
\usepackage{enumitem}

\usepackage{tabulary}
\usepackage{booktabs,longtable,tabularx}
\newcolumntype{L}{>{\raggedright\arraybackslash}X}
\usepackage{ltablex}
\usepackage{caption}
\usepackage{xcolor}
\usepackage{cuted}

\usepackage{selinput}
\SelectInputMappings{%
  aacute={á},
  ntilde={ñ},
  Euro={€}
}
\usepackage{mathtools}
\DeclarePairedDelimiter\abs{\lvert}{\rvert}%
\makeatletter
\let\oldabs\abs
\def\abs{\@ifstar{\oldabs}{\oldabs*}}

\usepackage[T1]{fontenc}
\usepackage{ae,aecompl}

\usepackage{newtxtext,newtxmath}

\title[GEOMAX compression]{GEOMAX: beyond linear compression for 3pt galaxy clustering statistics}

\author[D. Gualdi et al.]
{\parbox{\textwidth}{Davide Gualdi$^{1,2}$\thanks{Contact e-mail: \href{davide.gualdi.14@ucl.ac.uk}{dgualdi@icc.ub.edu}},
Héctor Gil-Marín$^{1,2}$,
Marc Manera$^{3,4}$,
Benjamin Joachimi$^{5}$,
}
\newauthor{
Ofer Lahav$^{5}$}
\vspace{0.4cm}\
\\
\parbox{\textwidth}
{
$^{1}$ICC, University of Barcelona, IEEC-UB, Mart\'i i Franqu\`es, 1, E08028 Barcelona, Spain\\
$^{2}$Institute of Space Studies of Catalonia (IEEC), E-08034 Barcelona, Spain\\
$^{3}$Institut de Física d’Altes Energies (IFAE), The Barcelona Institute of
Science and Technology, Campus UAB, 08193 Bellaterra (Barcelona) Spain\\
$^{4}$Centre for Mathematical Sciences, DAMTP, Cambridge University, Wilberforce Rd, Cambridge CB3 0W \\
$^{5}$Department of Physics and Astronomy, University College London, Gower Street, London WC1E 6BT, UK \\
}}

\date{XXXX-XXXX}

\pubyear{2019}

\begin{document}
\label{firstpage}
\pagerange{\pageref{firstpage}--\pageref{lastpage}}
\maketitle

\begin{abstract}
We present the GEOMAX algorithm and its Python implementation for a two-step compression of bispectrum measurements. The first step groups bispectra by the geometric properties of their arguments; the second step then maximises the Fisher information with respect to a chosen set of model parameters in each group. The algorithm only requires the derivatives of the data vector with respect to the parameters and a small number of mock data, producing an effective, non-linear compression. By applying GEOMAX to bispectrum monopole measurements from BOSS DR12 CMASS redshift-space galaxy clustering data, we reduce the $68\%$ credible intervals for the inferred parameters $\left(b_1,b_2,f,\sigma_8\right)$ by $\left(50.4\%,56.1\%,33.2\%,38.3\%\right)$ with respect to standard MCMC on the full
data vector. We run the analysis and comparison between compression methods over one hundred
galaxy mocks to test the statistical significance of the improvements. On average GEOMAX performs $\sim15\%$ better than geometrical or maximal linear compression alone and is consistent with being lossless. Given its flexibility, the GEOMAX approach has the potential to optimally exploit three-point statistics of various cosmological probes like weak lensing or line-intensity maps from current and future cosmological data-sets such as DESI, Euclid, PFS and SKA.

\end{abstract}




\begingroup
\let\clearpage\relax
\endgroup


\section{INTRODUCTION}
Recent applications of three-point (3pt) statistics to cosmological data-sets such as the one provided by the Sloan Digital Sky Survey \footnote{\url{http://www.sdss3.org/surveys/boss.php}} \citep{Eisenstein:2011sa}, have shown their potential in improving the current constraints on cosmological parameters \citep{Gil-Marin:2016wya,Slepian:2015hca}. For example the physics of baryonic acoustic oscillations has been investigated via 3pt statistics \citep{Slepian:2016kfz,Pearson:2017wtw} and new methods have been developed to better capture it \citep{Child:2018klv,Child:2018kec}, improving the results obtained via standard 2pt methods.

In preparation for future game-changing data-sets like DESI\footnote{\url{http://desi.lbl.gov}} \citep{Levi:2013gra}; Euclid \footnote{\url{http://sci.esa.int/euclid/}} \citep{Laureijs:2011gra}; PFS \footnote{\url{http://pfs.ipmu.jp}} \citep{Ellis:2012rn}; SKA\footnote{\url{https://www.skatelescope.org}} \citep{Bacon:2018dui} and LSST\footnote{\url{https://www.lsst.org/}} \citep{Abell:2009aa}, the scientific community has been working on improving the modelling of 3pt statistics and extending their applications.

\cite{Bertacca:2017dzm,Clarkson:2018dwn,DiDio:2018unb} studied the full sky angular galaxy bispectrum beyond the flat sky approximation including relativistic effects and corrections, \cite{Yamamoto:2016anp,Nan:2017oaq} and \cite{Sugiyama:2018yzo} proposed a complete multipole decomposition for the galaxy bispectrum.
\cite{Sabiu:2019kbh} proposed a technique to speed up the computation of 3pt and higher order statistics while \cite{DAmico:2019fhj} applied the effective field theory formalism for the galaxy bispectrum on data.

In the weak lensing field progress have been made since the first seminal works \citep{Takada:2003ef,Kilbinger:2005jy}, \cite{Rizzato:2018whp} investigated the information content of the bispectrum while \cite{Kayo:2012nm} forecasted parameters constraints using a covariance matrix with terms beyond the Gaussian approximation.
3pt statistic have also been applied to line intensity maps probes \citep{Hoffmann:2018clb,Beane:2018pmx,Schmit:2018rtf,Watkinson:2018efd}.

Studies on the bispectrum covariance and the bispectrum information content have also been realised in the recent past \citep{Colavincenzo:2018cgf,Yankelevich:2018uaz,Barreira:2019icq,Oddo:2019run}, while
\cite{Ruggeri:2017dda,Hahn:2019zob} and \cite{Coulton:2018ebd} proved that the bispectrum can also help with improving the sum of the neutrino masses constraints.
The accuracy of the bispectrum modelling was extended towards non-linear scales by including loop corrections \citep{Hashimoto:2017klo,Desjacques:2018pfv,Eggemeier:2018qae,Castiblanco:2018qsd}. Loop coorrections for the bispectrum 
\citep{Sefusatti:2009qh,Sefusatti:2011gt} 
must indeed be included if one wants to study primordial non-Gaussianity using mildly non-linear scales to lift the degeneracy with the other cosmological parameters \citep{Scoccimarro:2003wn,Bose:2018zpk,Karagiannis:2018jdt}.

Compression plays an essential role in the analysis of 3pt statistics. Indeed the difficulties to perform these analyses include the dimension of the associated data-vectors due to the limited number of simulations available to estimate the covariance matrix \citep{Hartlap:2006kj,Taylor:2014ota} and the computational challenge linked with the handling of large data-vectors. In particular when the estimators of these data-vectors are expressed in terms of multipole expansion, even including the quadrupole term becomes prohibitive. 
An alternative approach to compression would be to accurately model the analytic covariance matrix as done by \citep{Sugiyama:2019ike}, which helps with cross-validating the obtained results.

In a previous paper we introduced two methods achieving "maximal" compression for the redshift space galaxy bispectrum \citep{Gualdi:2017iey}, and tested them on bispectrum monopole measurements from BOSS DR12 data \citep{Gualdi:2018pyw}. 
These maximal compression methods transform the original data-vector into a new one with dimension equal to the number of parameters considered in the analysis. In order to work they require an approximate analytical expression of the original data-vector covariance matrix.

More recently in a second paper we presented the geometrical compression algorithm \citep{Gualdi:2019ybt} which is based on averaging the bispectra of wave-number triangle configurations having similar geometrical properties. 
Geometrical compression does not require an analytical expression for the covariance matrix but also does not take into account correlation between different triangle configurations.

In this work we present an algorithm along with its Python implementation\footnote{\url{https://github.com/davidegua/max_geo_compression.git}} that combines maximal and geometrical compression, accounting for the correlation between bispectra without needing an analytical expression for the covariance matrix. 
This is achieved in two steps. First we define triangle sets using the geometrical criteria. Secondly we apply the maximal compression separately to each of the triangles sets. 
A limited number of simulations for the maximal compression step is still needed. However, by maximally compressing each triangles set, the required number is much lower than what usually required to estimate the covariance matrix for the full data-vector.

We perform our analysis on measurements from both BOSS DR12 CMASS data \citep{Dawson:2012va} and on one hundred realisations of the relative set of mock data \citep{Kitaura:2015uqa}.

In the main analysis we sample the joint posterior distribution for four model parameters: matter-galaxy bias parameters $b_1$ and $b_2$, the growth rate $f$ and the amplitude of dark matter oscillations $\sigma_8$.
The main result of this work is a further improvement for the 68$\%$ credible intervals of the inferred parameters when using the joint data-vector including power spectrum (monopole and quadrupole) together with the bispectrum (monopole).

We run a series of tests to verify the added value of GEOMAX with respect to the previous methods.
First we consider alternative ways to first regroup triangles before applying the maximal compression step.
Second we run the analysis for alternative parameter sets. In one case we only add the local primordial non-Gaussianity parameter $f_\mathrm{NL}$. The bispectrum has indeed, especially for future data-set, the potential to produce constraints on $f_\mathrm{NL}$ of similar order \citep{Verde:1999ij,Sefusatti:2009qh,Jeong:2009vd} to the ones obtained by Planck \citep{Akrami:2019izv}.

In the second set we use $A_{\mathrm{s}}$, the amplitude of scalar perturbations and the matter density parameter $\Omega_{\mathrm{m}}$. These are of interest in order to obtain complementary late-time estimates on quantities very well constrained by cosmic microwave background (CMB) experiments \citep{Aghanim:2018eyx}.

The paper is organised as follows: in Section \ref{sec:previous_comp} we briefly recap the maximal and geometrical compression methods. 
Section \ref{sec:geo_max_compression} explains how to optimally combine maximal and geometrical compression methods, together with presenting the code structure. 
The analysis results for the main four parameters case are reported in Section \ref{sec:analysis_results} for both galaxy mocks and data. In Section \ref{sec:alternatives_to_geo} we consider two alternatives to the geometrical compression step while in Section \ref{sec:alternative_par_sets} we repeat the analysis for two larger parameter sets.  We conclude in Section \ref{sec:conclusions}. In Appendix \ref{sec:appendix_models} the analytical expression for the used data vectors are reported while in Appendix \ref{sec:appendix_png} we re-derive the primordial non-Gaussianity leading correction terms for the power spectrum and bispectrum.
Before starting with the main part of the paper we summarise below the analysis setup.

\subsection{Analysis setup}
\label{sec:dataset}
For a fair comparison with our previous papers we once more apply the compression to the measurements of the galaxy bispectrum monopole from the DR12 CMASS sample ($0.43\leq z \leq 0.70$) of the Baryon Oscillation Spectroscopic Survey (BOSS, \citealp{Dawson:2012va}) which is part of the Sloan Digital Sky Survey III \citep{Eisenstein:2011sa}.

We use 1400 realisations of the MultiDark Patchy galaxy catalogues for the BOSS DR12 data-set by \cite{Kitaura:2015uqa}. These mocks were realised having a fiducial cosmology with parameters  $\Omega_{\Lambda}(z=0) = 0.693$, $\Omega_{\mathrm{m}}(z=0) = 0.307$, $\Omega_{\mathrm{b}}(z=0) = 0.048$, $\sigma_8=0.829$, $n_{\mathrm{s}} = 0.96$, $h_0 = 0.678$.

The data-vector used for the parameter constraints analysis is given by joining the galaxy tree-level power spectrum monopole and quadrupole measurements to the bispectrum monopole ones. The analytical expressions are given in Appendix \ref{sec:appendix_models}.
For the power spectrum part we use a bin size of $\Delta k = 0.01 h/$Mpc while for the bispectrum we use two different sizes proportional to the fundamental frequency $k_f = \frac{(2\pi)^3}{V_\mathrm{s}}$.  $V_\mathrm{s} = (3500\,\mathrm{Mpc}/h)^3$ is the survey volume for the cubic box used to generate the mock catalogues. For the bispectrum we then consider the two cases $\Delta k_{6,2} = 6,2 \times k_f$ corresponding to 116 and 2734 triangle configurations, respectively.

As done in the previous works \citep{Gualdi:2019ybt,Gualdi:2018pyw} we use $0.03\,\ h/\mathrm{Mpc}\leq k \leq0.09 h/\mathrm{Mpc}$ for the power spectrum terms and $0.02\, h/\mathrm{Mpc}\leq k \leq0.12 h/\mathrm{Mpc}$ for the bispectrum monopole. The bispectrum has a larger $k$-range because we adopted the effective kernel calibrated on simulations used also for the BOSS DR11 and DR12 analysis \citep{Gil-Marin:2014sta,Gil-Marin:2016wya} which allows to safely extend the analysis to mildly non-linear scales \citep{GilMarin:2011ik}.

All the MCMC samplings (both on original and compressed data-vectors) have been run with the same settings as in previous works \citep{Gualdi:2018pyw,Gualdi:2019ybt}.

Finally we use a flat $\Lambda$CDM cosmology to compute the linear matter power spectrum, with parameters close to the results from the Planck analysis \citep{Akrami:2019izv}, in particular $\Omega_{\mathrm{m}}(z=0) = 0.31$, $\Omega_{\mathrm{b}}(z=0) = 0.049$, $A_{\mathrm{\mathrm{s}}}=2.21\times 10^{-9}$, $n_{\mathrm{s}} = 0.9624$, $h_0 = 0.6711$ and $\sum m_{\nu} = 0.06$ eV.

\section{Previous compressing methods}
\label{sec:previous_comp}

\subsection{Maximal compression}
\label{sec:maximal_compression}
Maximal linear compression derives from the 
MOPED method \citep{Heavens:1999am}, which compresses the original data-vector by extending to the multiple parameters case the algorithm introduced by \cite{Tegmark:1996bz}.
As a result, an originally arbitrarily large data-vector $\bm{x}$ can be transformed into a much shorter one $\bm{y}$ having dimension equal to the number of model parameters considered in the analysis.
This is achieved by taking the scalar product of $\bm{x}$ with a set of weights $\bm{b}_i$ for each of the model parameters $\theta_i$:

\begin{eqnarray}
y_i\, = \,\langle\bm{x}\rangle_{,i}^\intercal\cdot\,\mathrm{\textbf{Cov}}_{\bm{x}}^{-1}\cdot\,\bm{x} \,\equiv\, \bm{b}^\intercal_i\cdot\,\bm{x},
\label{eq:max_com}
\end{eqnarray}

\noindent where $\mathrm{\textbf{Cov}}$ is the covariance matrix for the original data-vector $\bm{x}$ while $\langle\bm{x}\rangle_{,i}$ are the derivatives of the mean of the modelled data-vector with respect to the model parameters $\theta_i$. The maximal compression method presented in \cite{Gualdi:2017iey} that we consider in this paper consists in running an MCMC sampling on the compressed bispectrum data-vector.

\subsection{Geometrical compression}
\label{sec:geo_compression}

The main idea in \cite{Gualdi:2019ybt} is to group together into new bins the triangles with similar geometrical properties. Each of these bins will correspond to an element of the compressed data-vector. The value of the data-vector's element is then given by the average bispectra of the triangle configurations belonging to the same bin.

The standard parametrisation of each triangle configuration is given in terms of the three sides $\left(k_1,k_2,k_3\right)$.
The new parameters are chosen using the physical intuition regarding which quantities most influence the bispectrum value. 
These are:

\begin{itemize}
    \item the square root of the triangle's area: $\aleph$ ("aleph");
    \item the cosine of the largest internal angle, $\daleth = \cos\psi_{\mathrm{max}}$ ("daleth");
    \item the ratio between the cosines of the intermediate and smallest angles, $\gimel = \cos\psi_{\mathrm{int}}/\cos\psi_{\mathrm{min}}$ ( "gimel").
\end{itemize}

\noindent We rewrite the galaxy bispectrum monopole data-vector as a function of the three new variables $(\aleph,\daleth,\gimel)$:

\begin{eqnarray}
\mathrm{B}^{(0)}_{\mathrm{g}}\left(k_1,k_2,k_3\right) \quad \Longrightarrow \quad \mathrm{B}^{(0)}_{\mathrm{g}}\left(\aleph,\daleth,\gimel\right) . 
\end{eqnarray}

\noindent 
By choosing large enough bins for the new variables, triangles with similar $(\aleph,\daleth,\gimel)$ coordinates will be grouped together. 

The new data-vector $\bm{g}$ is then obtained by averaging over all the bispectra in the new bins defined by different sets of the coordinates $(\aleph,\daleth,\gimel)$:

\begin{eqnarray}
\label{eq:algorithm}
g_k(\aleph, \daleth, \gimel)_k = \dfrac{1}{N^{\mathrm{tr.}}_k}
\sum^{N^{\mathrm{tr.}}_k}_{j\,:\,(k_1,k_2,k_3)_j\in(\aleph, \daleth, \gimel)_k}
\mathrm{B}^{(0)}_{\mathrm{g}}(k_1,k_2,k_3)_j\,.
\end{eqnarray}

\noindent Each new data-vector element has been normalised dividing by $N^{\mathrm{tr.}}_k$, the number of triangles belonging to the same bin obtained from a particular choice of $(\aleph,\daleth,\gimel)_k$.


\setlength{\tabcolsep}{4pt}
\begin{table}
	\centering
	  \caption[Improvement in parameter constraints]{ The improvements in parameter constraints shown are the relative change of the $68\%$ credible intervals for the $\Delta k_2$  $k$-binning case with respect to the $\Delta k_6$. While maximal and geometrical methods achieve similar results, their combination, GEOMAX, on average performs better by $\sim15\%$ when applied to BOSS DR12 CMASS data.}
	\label{tab:improvements_4par}
  \begin{tabular}{ *{5 }{c}}
\toprule

& $\Delta\theta^{\mathrm{mc}}_{\Delta k_6}$ 
& \multicolumn{3}{c}{$\dfrac{ \Delta\theta^{\mathrm{mc}}_{\Delta k_6}-\Delta\theta^{\mathrm{comp.}}_{\Delta k_2}}{\Delta\theta^{\mathrm{mc}}_{\Delta k_6}}\;\left[\%\right]$}  \\
\cmidrule(lr){2-2}\cmidrule(lr){3-5}
        &  MCMC   & MAX  & GEO & GEOMAX \\
\cmidrule(lr){2-2}\cmidrule(lr){3-5}
        &  $N_{\mathrm{tr}}=116$   & $N_{\mathrm{el.}}=4$ &$N_{\mathrm{g}}=116$ &$N_{\mathrm{g}}=116$ \\
\cmidrule(lr){2-2}\cmidrule(lr){3-5}
$\Delta b_1 $     & 0.22 & \colorbox{-red!75!green}{36.1} &  \colorbox{yellow!70}{28.6} & \colorbox{red!75}{50.4}\\
$\Delta b_2 $     & 0.40 & \colorbox{-red!75!green}{45.5} &  \colorbox{yellow!70}{36.5} & \colorbox{red!75}{56.1}\\
$\Delta f   $     & 0.08 & \colorbox{-red!75!green}{23.9} &  \colorbox{yellow!70}{18.5} & \colorbox{red!75}{33.2}\\
$\Delta \sigma_8$ & 0.04 & \colorbox{-red!75!green}{21.6} &  \colorbox{yellow!70}{15.1} & \colorbox{red!75}{38.3}\\
\cmidrule(lr){1-5}
\multicolumn{2}{c}{$\Big\langle\dfrac{
\Delta\theta^{\mathrm{mc}}_{\Delta k_6}-\Delta\theta^{\mathrm{comp.}}_{\Delta k_2}}{\Delta\theta^{\mathrm{mc}}_{\Delta k_6}}\;\left[\%\right]\Big\rangle$} &31.8 & 24.7 & 44.5\\

\bottomrule
    \end{tabular}
\end{table}

\begin{figure*}
  \begin{subfigure}[t]{0.49\textwidth}
    \includegraphics[width=\textwidth]
    {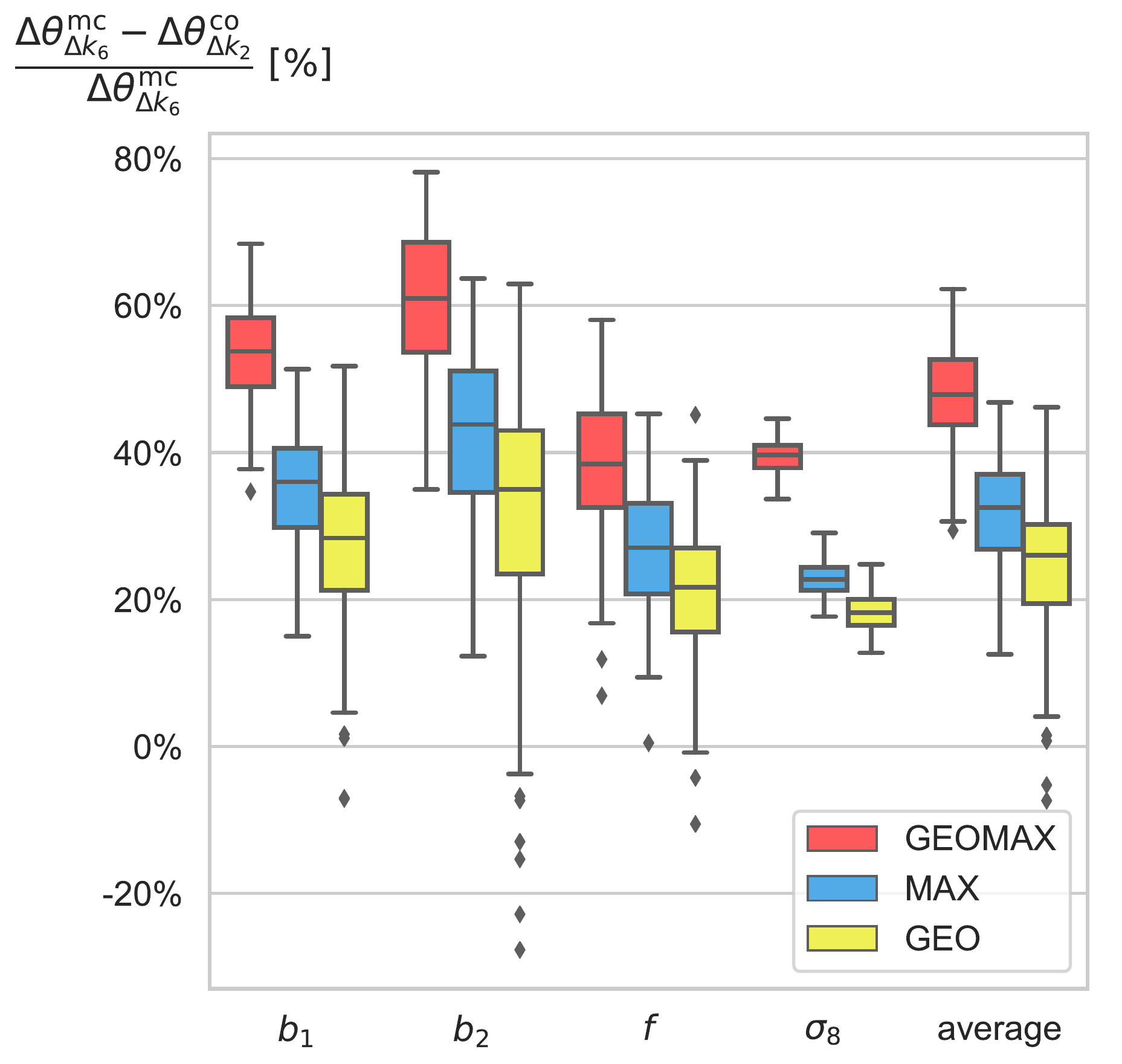}
    \caption{Mocks}
  \end{subfigure}
  \begin{subfigure}[t]{0.49\textwidth}
    \includegraphics[width=\textwidth]
    {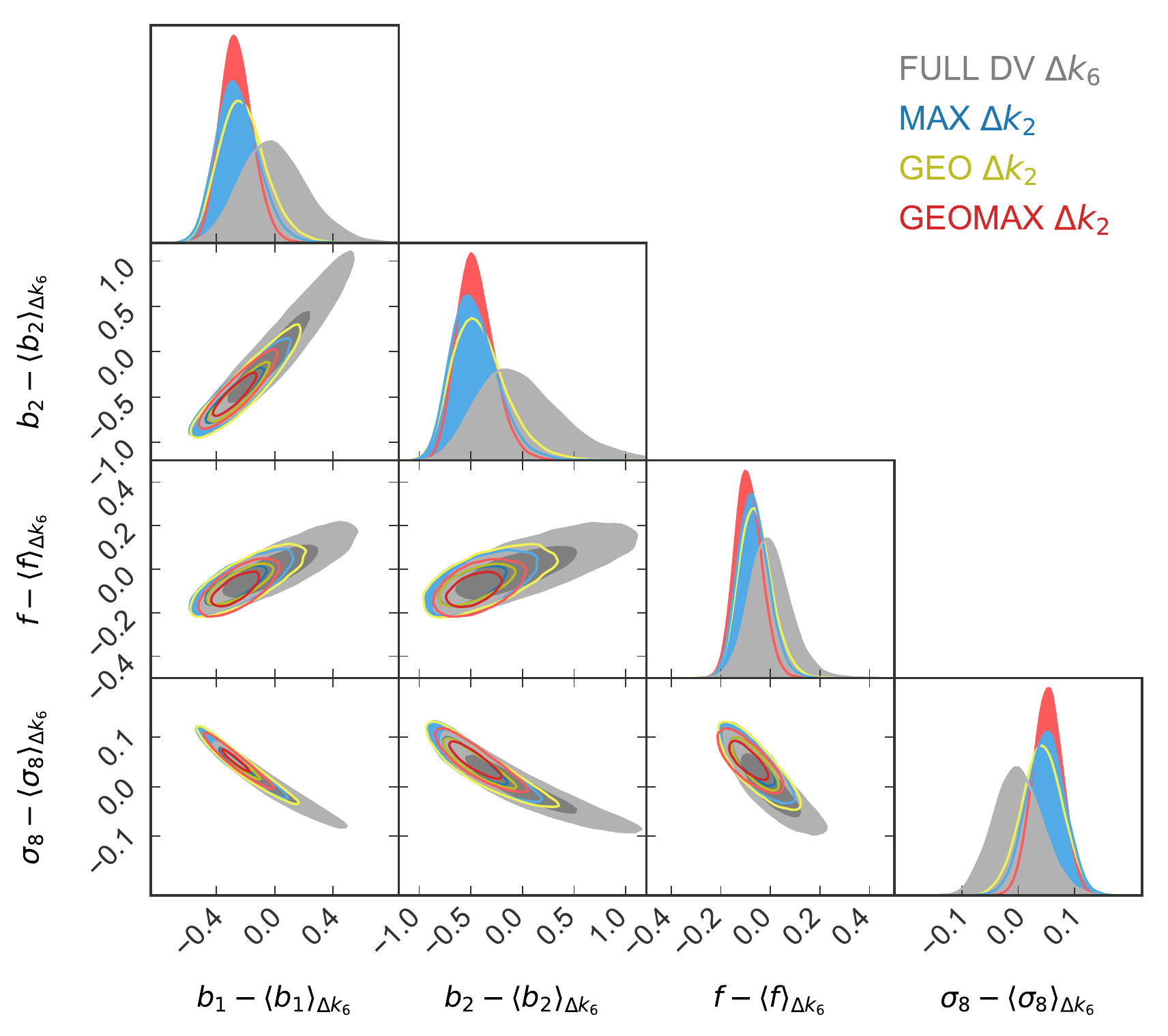}
    \caption{Data}
  \end{subfigure}
      \caption{
     Joint data-vector $\left[\mathrm{P}^{(0)}_{\mathrm{g}},\mathrm{P}^{(2)}_{\mathrm{g}},\mathrm{B}^{(0)}_{\mathrm{g}}\right]$ posteriors: four parameters case \newline
\textbf{a)}
  The above boxplot summarises the statistical distribution of the parameter constraints improvements for the new geometric-maximal method (GEOMAX) and the two individual maximal (MAX) and geometrical (GEO) ones. The coloured boxes show the central quartiles of each distribution while the individual dots are automatically considered outliers by the plotting routine. Each boxplot is obtained comparing the different methods (compressing 2734 triangles, $\Delta k_2$ case) with the MCMC on the full data-vector (116 triangles, $\Delta k_6$ case) for 100 realisations of the Patchy Mocks \citep{Kitaura:2015uqa}.
  The last column showing the average improvement has the purpose to explain the reason for which certain mocks have some of the parameters with negative improvements (in particular for the GEO compression). The always positive average improvement (beside for two mocks out of one hundred) shows that the individual negative ones are a statistical fluctuation due the above average improvements for the other parameters.
    \newline
\textbf{b)} 
    Compression performance: 2-D  $68\%$ and $95\%$ credible regions are shown for the standard MCMC sampling on the full data vector using the triangles from the $\Delta k_6$ case (116 triangles, MCMC). For the $\Delta k_2$ case (2734 triangles) are shown the contours obtained by compressing the bispectrum part of the data-vector using maximal (\citealp{Gualdi:2017iey,Gualdi:2018pyw}, MAX), geometrical (\citealp{Gualdi:2019ybt}, GEO) and enhanced geometrical compression methods (GEOMAX). The combination of geometrical and maximal compressing methods (GEOMAX) further improves the parameters constraints as quantitatively described in Table \ref{tab:improvements_4par}. These marginalised posterior distributions have been derived using measurements from BOSS DR12 CMASS data.
    We subtracted to all the distributions the central value obtained for the $\Delta k_6$ standard MCMC case. This because our goal is to test whether we would observe on data, which usually include unknown systematics, the same improvements statistically observed on galaxy mock catalogues.
}
    \label{fig:data_mocks_4p}
\end{figure*}

\section{ENHANCED GEOMETRICAL COMPRESSION}
\label{sec:geo_max_compression}
We combine geometrical and maximal compression, labelling this new method GEOMAX, in the following way. First we regroup triangles as described in Section \ref{sec:geo_compression}. Secondly, instead of averaging over all the bispectra of the triangle configurations belonging to each bin, we separately apply the maximal compression to each bin. Therefore for every bin defined by a set of new coordinates $(\aleph, \daleth, \gimel)_k$ we will obtain, for each of the model parameters $\theta_i$, a compressed data-vector element $g_{ik}^{\mathrm{opt.}}$ defined by 

\begin{eqnarray}
g_{ik}^{\mathrm{opt.}}(\aleph, \daleth, \gimel)_k =  \bm{b}_{ik}\cdot\,
\bm{\mathrm{B}}^{(0)}_{\mathrm{g},k}\,.
\end{eqnarray}

\noindent $\bm{\mathrm{B}}^{(0)}_{\mathrm{g},k}$ is the reduced data-vector formed by the bispectra $\mathrm{B}^{(0)}_{\mathrm{g}}(k_1,k_2,k_3)_j$ of the triangles belonging to the bin defined by $(\aleph, \daleth, \gimel)_k$, such that ${j\,:\,(k_1,k_2,k_3)_j\in(\aleph, \daleth, \gimel)_k}$. The weights vector $\bm{b}_{ik}$ for the $k$-bin, according to the definition given in Eq. \ref{eq:max_com}, is given by,

\begin{eqnarray}
\label{eq:weights_max_geo}
\bm{b}_{ik}\, = \,\left(\dfrac{\partial \langle\bm{\mathrm{B}}^{(0)}_{\mathrm{g},k}\rangle}{\partial \theta_i}\right)^\intercal\cdot\,\bm{\mathrm{Cov}}^{-1}_{\bm{\mathrm{B}}^{(0)}_{\mathrm{g},k}}\,,
\end{eqnarray}

\noindent where $\bm{\mathrm{Cov}}^{-1}_{\bm{\mathrm{B}}^{(0)}_{\mathrm{g},k}}$ is the covariance matrix for the reduced data-vector $\bm{\mathrm{B}}^{(0)}_{\mathrm{g},k}$ of the bin $k$ computed using the available simulations or galaxy mock catalogues. In our case we used 1400 realisation of the MultiDark Patchy BOSS DR12 mocks. We adopt the conservative approach of ensuring that the number of triangle configurations belonging to each new bin is less than half the number of available mocks. 
This is to reduce to a reasonable level the bias induced by estimating the covariance matrix from a limited number of realisations \citep {Hartlap:2006kj}. In any case this bias would be a constant factor and therefore not affecting the compression weights.
We can then assume that $\bm{\mathrm{Cov}}^{-1}_{\bm{\mathrm{B}}^{(0)}_{\mathrm{g},k}}$ is a good approximation of the covariance matrix of the reduced data-vector bin $\bm{\mathrm{B}}^{(0)}_{\mathrm{g},k}$. 
Therefore in order to estimate the covariance matrix needed to maximally compress the largest bin (which includes $\sim 500$ triangle configurations) we need at least $\sim1000$ mock catalagoues. This requirement is more than five times lower than the number of mocks needed when the full data-vector (2734 triangle configurations) is considered ($\simeq 5500$ simulations).

In the algorithm selecting the best bin number settings it is possible to impose a maximum number of triangle configurations per bin. In this way one can arbitrarily reduce the number of simulations required to estimate the covariance matrix for each bin. However excessively reducing the number of maximum triangles per bin would also decrease the performance of the compression. In the same way, increasing the maximum number of triangle configurations per bin up to the number of simulations available would decrease the overall performance. This is because the covariance matrices used for the maximal compression step would no longer be accurate enough, hence reducing the efficacy of the compression weights.

\subsection{Updated optimal binning choice selection criteria}
\label{sec:updated_sel_crit}

In order to choose the number of bins for the new data-vector, or in other words, the triplets of $(\aleph,\daleth,\gimel)_k$ defining the elements of the new data-vector, in \cite{Gualdi:2019ybt} we estimated the sensitivity of the potential compressed data-vector with respect to the considered cosmological parameters. This was done by computing a summary statistic for each possible choice of bins numbers. This number is obtained by averaging together the derivatives of the compressed data-vector $\bm{g}$. For the geometrical compression, it was fundamental to divide each compressed data-vector element's derivative  by the number of triangle configurations corresponding to the bin $k$ before averaging: 

\begin{eqnarray}
\mathrm{S}_{ij} \equiv \sum_{k=1}^{N_{\bm{g}}(n_{\aleph}, n_{\daleth}, n_{\gimel})_j} \,\dfrac{1}{N_k^\mathrm{tr.}}\, \abs{\dfrac{\partial g_k}{\partial\theta_i}},
\end{eqnarray}

\noindent where $\mathrm{S}_{ij}$ is an estimator of the sensitivity of $\bm{g}$ when varying the model parameter $\theta_i$, defined for a particular choice of number of bins $(n_{\aleph}, n_{\daleth}, n_{\gimel})_j$.

For GEOMAX this no longer works since the new data-vector elements are derived from a linear combination of the original bispectra, where the weights are given by the maximal compression applied to each bin as shown in Eq. \ref{eq:weights_max_geo}. 
Therefore we need to define a new summary statistic which normalises each compressed data-vector's element derivative dividing by the sum of the weights used to compute it:

\begin{align}
\label{eq:new_summary}
\mathrm{S}^{\mathrm{gm}}_{ij} &\equiv \sum_{k=1}^{N_{\bm{g}}(n_{\aleph}, n_{\daleth}, n_{\gimel})_j}\,\sum_{\ell=1}^{N_{\mathrm{par}}} \,\left(\sum_
{m=1}^{N_k^\mathrm{tr.}}b_{\ell k}^m\right)^{-1}\, 
\dfrac{\partial g_{\ell k}}{\partial\theta_i}
\notag \\
&=
\sum_{k=1}^{N_{\bm{g}}(n_{\aleph}, n_{\daleth}, n_{\gimel})_j}\,\sum_{\ell=1}^{N_{\mathrm{par}}} \,\left(\sum_
{m=1}^{N_k^\mathrm{tr.}}b_{\ell k}^m\right)^{-1}\,
\bm{b}_{\ell k}\cdot\,\dfrac{\partial \bm{B}^{(0)}_k }{\partial \theta_i}
,
\end{align}

\noindent where the sum over $k$ accounts for all the elements of the compressed data-vector. The sum over $\ell$ covers the number of linear combinations (equal to the number of model parameters) obtained from each triangle configurations group defined by a set $(\aleph,\daleth,\gimel)_k$. The self-contained sum over $m$ inside the curved brackets acts as a normalisation factor specific for each of the compressed data-vector derivative's elements.

We can then proceed as in the case of the geometrical compression where a single number can be obtained by:

\begin{eqnarray}
\label{eq:final_bin_estimator}
\bar{\mathrm{s}}^{\mathrm{gm}}_{j} \equiv \sum_{i = 1}^{N_{\bm{\theta}}}\, \mathrm{s}^{\mathrm{gm}}_{ij}
=
\sum_{i = 1}^{N_{\bm{\theta}}}\,
\dfrac{\mathrm{S}^{\mathrm{gm}}_{ij}}{\mathrm{max}\left[\mathrm{S}^{\mathrm{gm}}_{ij}\right]_{\forall j}},
\end{eqnarray}

\noindent Again we choose the set of $(n_{\aleph}, n_{\daleth}, n_{\gimel})_j$ which maximises $\bar{s}^{\mathrm{gm}}_j$.


\begin{figure}%
    \centering
    \includegraphics[width=0.45\textwidth]
    {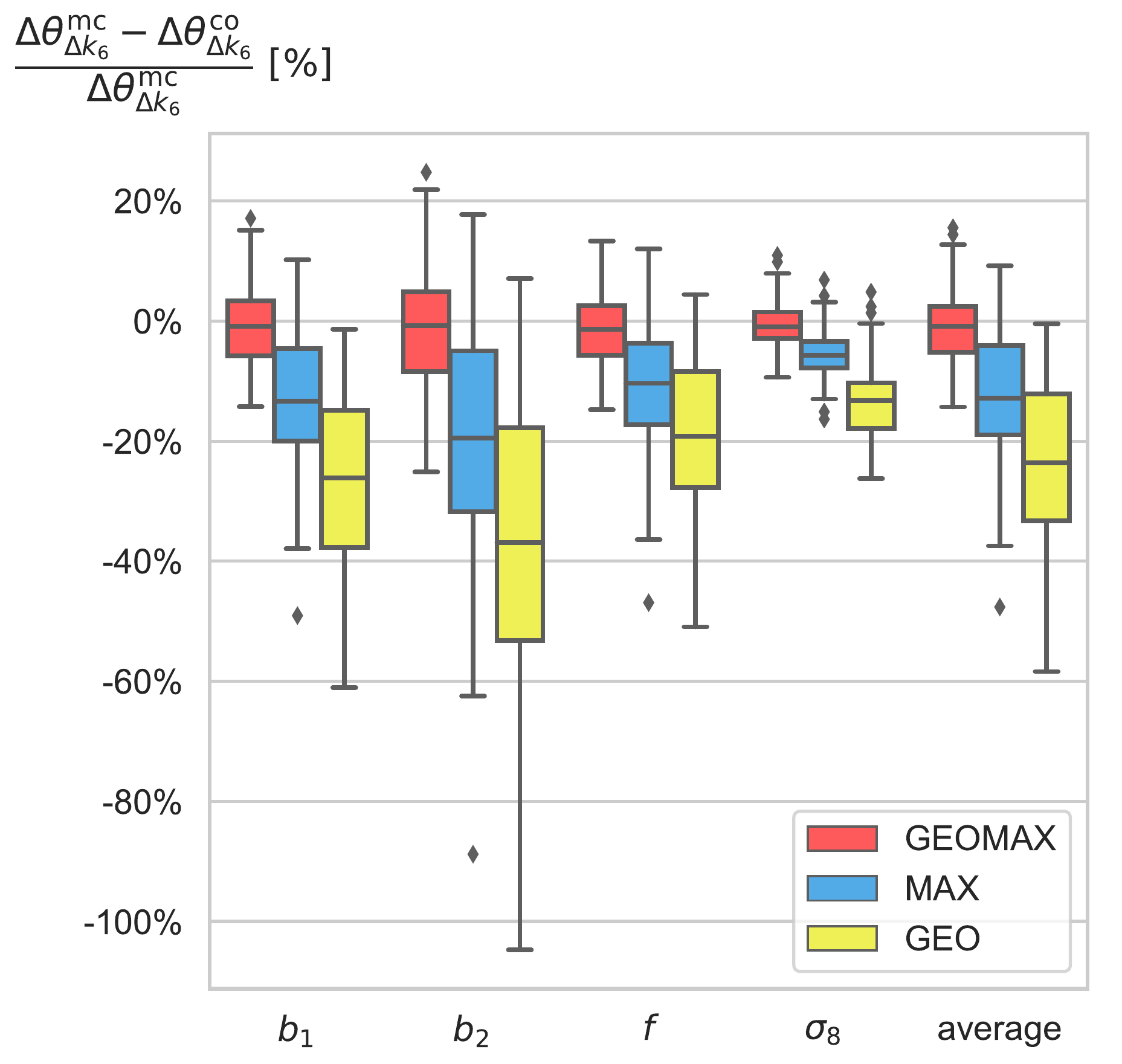}
    \caption{Loss of information test: using the bispectra values of the 116 triangle configurations corresponding to the $\Delta k_6$ binning case, we check whether the  compressed data-vectors retain the same information of the original full one. We show for the three different methods, enhanced geometrical c. (GEOMAX), maximal c. (MAX) and geometrical c. (GEO), the ratio between the width of the 1D $68\%$ credible regions with the ones obtained running an MCMC sampling on the full data-vector. The distributions for the different parameters are obtained by computing the ratio for 100 galaxy catalogues measurements. From the above box-plots we see that only GEOMAX is statistically consistent with zero loss of information.
}
    \label{fig:loss_info4p}
\end{figure}

\subsection{Code structure}
\label{sec:code_presentation}
In order to derive the compressing function, three ingredients are required:
\begin{itemize}[leftmargin=+.2in]
    \item the triangle configurations in terms of the sides length ($k_1$,$k_2$,$k_3$);
    \item the derivatives of the 3pt data-vector with respect to the model parameters;
    \item measurements of the 3pt data-vector from the available simulations.
\end{itemize}

\noindent Moreover one can set the maximum number of triangles per bin and the range to check for the number of bins for each of the geometrical parameters.

The code\footnote{\url{https://github.com/davidegua/max_geo_compression.git}} consists in creating an object with several functions needed to find the optimal compression as well as to convert the analytical model and measurements of the data-vector into their compressed form.
First of all, the code computes the optimal binning in terms of geometrical compression, using the updated selection criteria presented in the previous section. Afterwards, bins with less triangles than the number of model parameters are merged together into a single bin. Finally the weights to maximally compress each group of triangles are obtained using Eq. \ref{eq:max_com} where the covariance matrix is estimated using the available simulations (whose number has to be larger than the number of triangles in each group).

The created object contains all the information required to convert the data-vectors and the measurements in their compressed form. This can be done by using the object's methods. For example in our analysis pipeline this is done at every step of MCMC sampling: we first compute the full data-vector which is subsequently compressed (together with the covariance matrix) before the likelihood evaluation.


\begin{figure*}%
    \centering
    \includegraphics[width=0.9\textwidth]
    {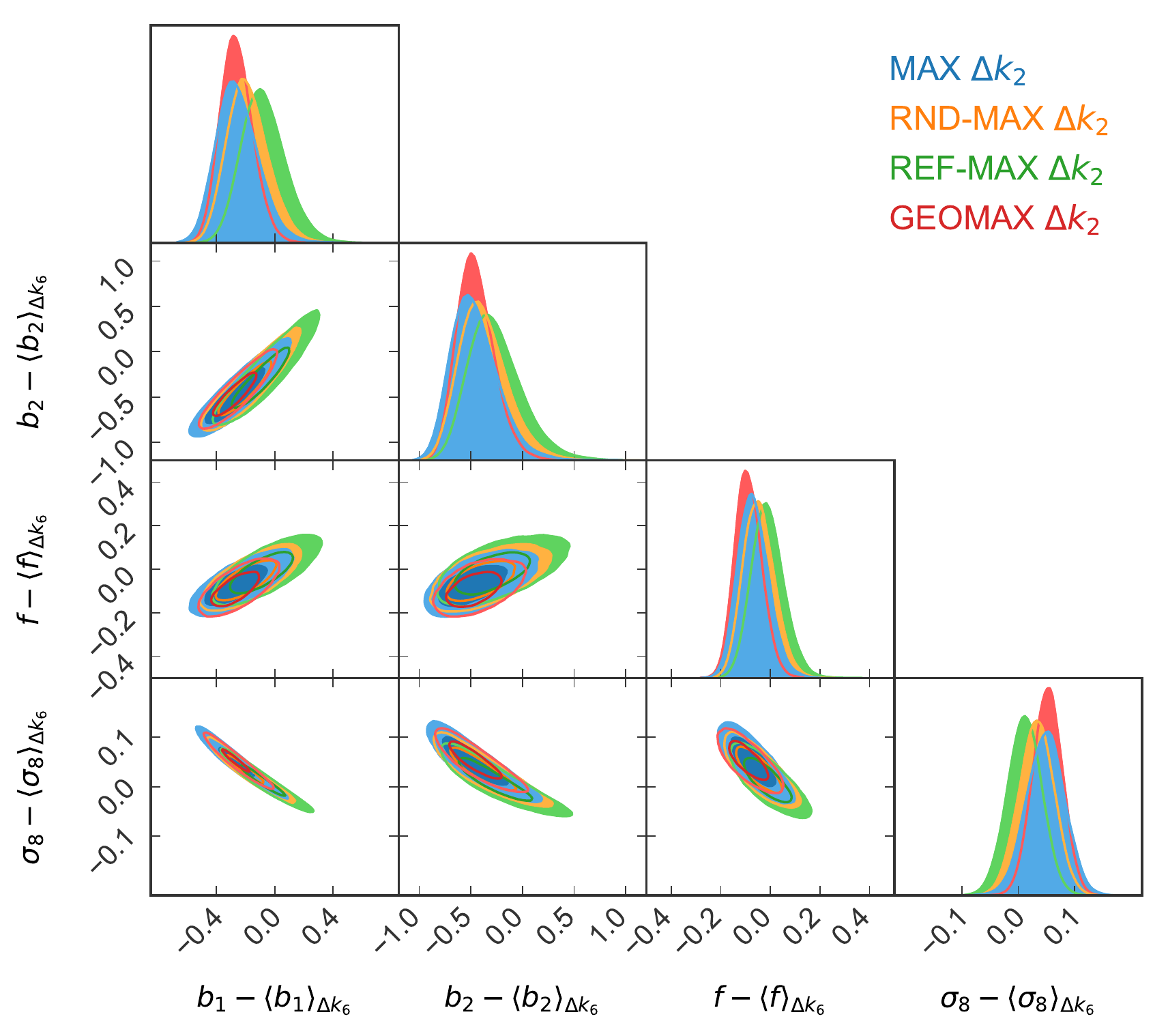}
    \caption{
    Compression performance of the alternatives to the geometrical compression step: 2-D  $68\%$ and $95\%$ credible regions are shown for data-vectors obtained by grouping together the 2734 $\Delta k_2$ triangles in the two alternative ways described in Section \ref{sec:alternatives_to_geo}. RND-MAX corresponds to the random assignation (triangles equally distributed) to 29 bins before the maximal compression. For REF-MAX, the triangles are associated to the bin whose "reference" triangle is closer in terms of perimeter. Comparing the contours of RND-MAX and REF-MAX with the ones for GEOMAX we can see the importance of the geometrical compression step. Without it, there is no significant improvement with respect maximal compression (MAX) alone.
    The different 1D marginalised posterior distributions shifts with respect to the central values obtained by the MCMC sampling are due to the different reduction of the parameter space degeneracy achieved by the methods shown. The smaller is the improvement with respect to the MCMC constraints, the smaller is the shift. This was previously discussed in \citep{Gualdi:2018pyw,Gualdi:2019ybt}.
    }
    \label{fig:alternatives_to_geo}
\end{figure*}
\section{RESULTS ANALYSIS}
\label{sec:analysis_results}

We compare the enhanced geometrical compression (GEOMAX) with the previous works including the MCMC on the full data-vector obtained by considering the same $k$-bins size ($\Delta k_6$), as in the BOSS analysis \citep{Gil-Marin:2014sta,Gil-Marin:2016wya}, which for our $k$-range choice corresponds to 116 triangle configurations. For all the compression methods we use 2734 triangle configurations obtained by choosing a three times smaller $k$-bin size ($\Delta k_2$). We cannot use the full data-vector with 2734 triangle configurations since there are not enough galaxy mock catalogues available to numerically estimate 
an invertible covariance matrix \citep{Hartlap:2006kj}. Maximal compression (MAX) returns a compressed data-vector with a number of elements equal to the number of model parameters. For the geometrical (GEO) and enhanced geometrical (GEOMAX) compression methods we impose, 
for a fair comparison with standard MCMC, a maximum number of bins equal to the dimension of the full data-vector (116 triangles) plus one. The two algorithms return a compressed data-vector with dimension equal to 116 (GEO) and 115 (GEOMAX), respectively .


\begin{figure*}
  \begin{subfigure}[t]{0.49\textwidth}
    \includegraphics[width=\textwidth]
    {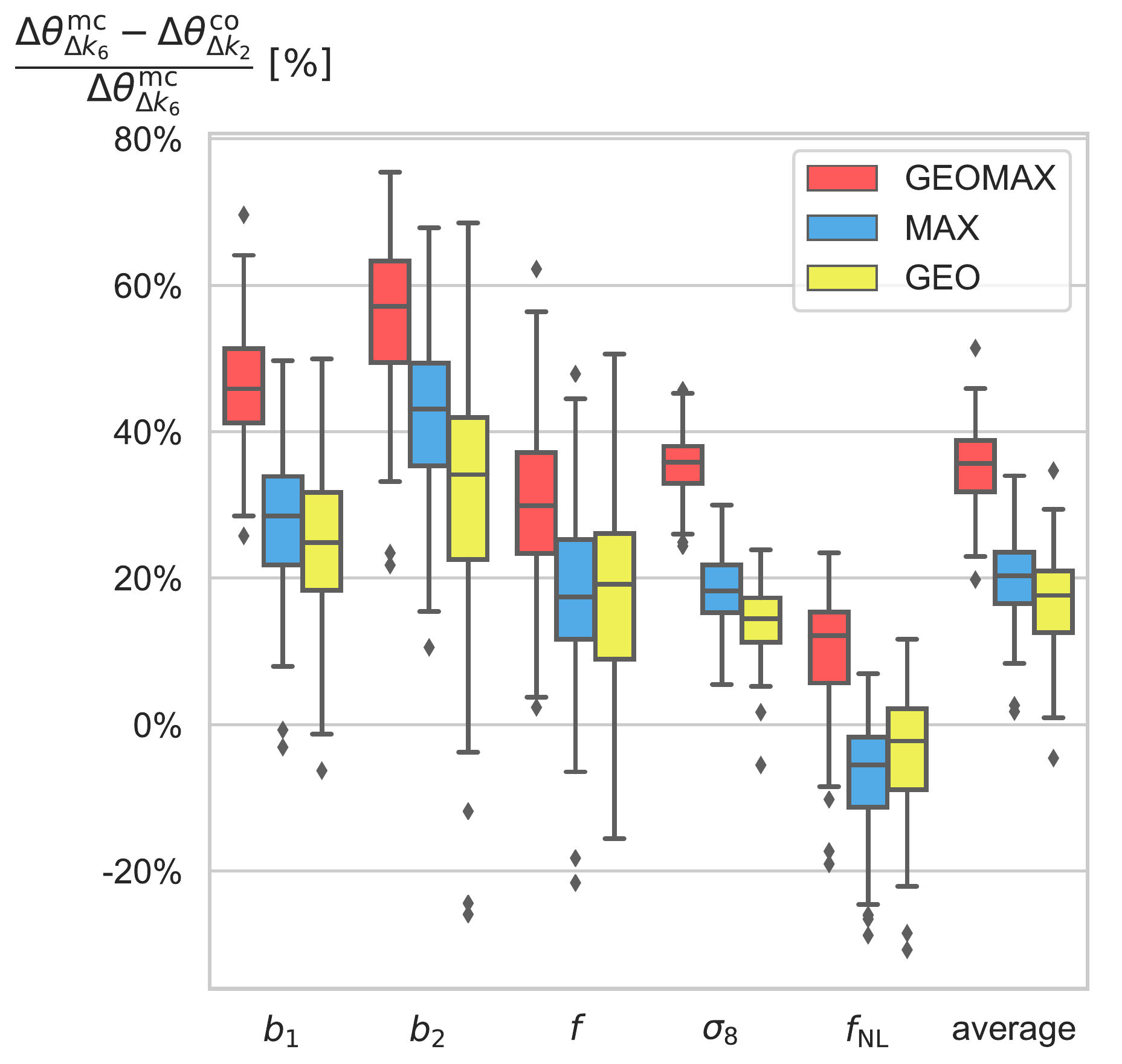}
    \caption{Mocks}
  \end{subfigure}
  \begin{subfigure}[t]{0.49\textwidth}
    \includegraphics[width=\textwidth]
    {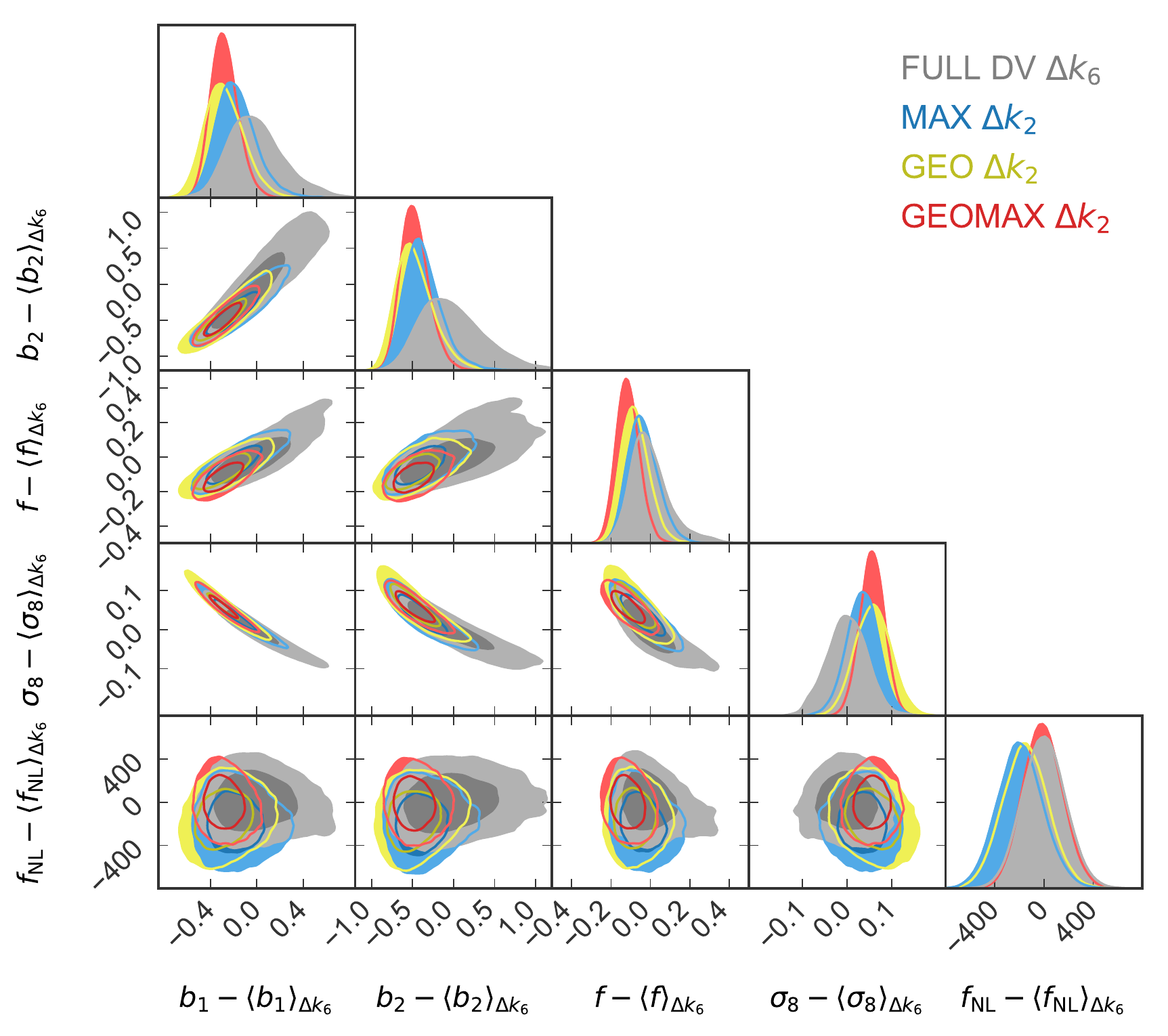}
    \caption{Data}
  \end{subfigure}
  \caption{Joint data-vector $\left[\mathrm{P}^{(0)}_{\mathrm{g}},\mathrm{P}^{(2)}_{\mathrm{g}},\mathrm{B}^{(0)}_{\mathrm{g}}\right]$ posteriors: five-parameter case including local primordial non-Gaussianity. Same as Figure \ref{fig:data_mocks_4p} when considering the additional parameter $f_\mathrm{NL}$.}
\label{fig:data_mocks_5p}
\end{figure*}


\subsection{Statistical significance from galaxy mocks}
\label{sec:stat_significance}
We validate our method by studying the distribution of the improvements, over the 1D $68\%$ credible regions on the parameter constraints, by both running MCMC samplings on the full data-vector ($\Delta k_6$) and on the compressed one for MAX, GEO and GEOMAX methods ($\Delta k_2$) on 100 realisations of the Patchy Mocks. 

In the left panel of Figure \ref{fig:data_mocks_4p} we can appreciate how GEOMAX outperforms, considering the means of the distributions, the MAX and GEO methods by obtaining approximately on average $15\%$ tighter constraints. With respect to the standard MCMC, the 1D $68\%$ credible regions are $40-60\%$ tighter when using GEOMAX. 
Moreover the improvements values scatter for GEOMAX is on average smaller than the ones for both MAX and GEO methods.

The column showing the average in the left panel of Figure \ref{fig:data_mocks_4p} explains why for certain mocks some parameters have negative improvements.
Since the average is always positive we can explain the individual negative improvements as a statistical fluctuation due to the other parameters above average improvements. In Appendix \ref{sec:outliers} we test for one of these mocks that the negative improvements for some parameters are not due to the choice of fiducial cosmology used to derive the compression.


For the different methods we also test the loss of information associated with the compression of the data-vector.
In Figure \ref{fig:loss_info4p} we show once more the ratio of the 1D $\%$ credible regions, however in this case we compress the same 116 triangle configurations long data-vector used for the standard MCMC ($\Delta k_6$). For GEOMAX and GEO methods we set a maximum number of elements for the compressed data-vector equal to 60. We chose this threshold since for the GEOMAX method it corresponds (in the case of four parameters) to a maximum number of bins for the geometrical step equal to 15. Figure \ref{fig:loss_info4p} highlights that, even in this "few-bins" possible scenario for the geometrical step, GEOMAX compression is statistically consistent with zero loss of information with respect to the MCMC on the full data-vector. The geometrical compression suffers more from the few bins available since in each bin the bispectra values are averaged, whilst in GEOMAX they are weighted by the coefficients given by the maximal compression step given by Eq. \ref{eq:weights_max_geo}.

\subsection{Test on BOSS DR12 CMASS data}
We apply our method on data to test whether we find a performance similar to what statistically observed using the measurements from the galaxy mocks. 
Qualitative results for the data can be observed in the right panel of Figure \ref{fig:data_mocks_4p}.
Since we are mainly interested in the improvements on the parameters constraints, we present our results with the central value of each parameter obtained through standard MCMC sampling subtracted.
We will perform a full parameter constraints analysis, including also Finger of God and Alcock-Paczynski effects, in a following paper, extending the modelling in order to include smaller scales.

In Table \ref{tab:improvements_4par} we can see the improvements obtained by the compression methods for each of the parameter constraints derived using the MCMC on the full data-vector, maximal compression (MAX), geometrical compression (GEO) and enhanced geometrical compression (GEOMAX).

For what concerns the model parameters 1D $68\%$ credible intervals, GEOMAX performs on data on average $45.5\%$ better than standard MCMC and improves the results obtained by maximal and geometrical compression by approximately $\sim 15\%$. This shows very good agreement with the improvements observed in the case of the galaxy catalogues shown in the left panel of Figure \ref{fig:data_mocks_4p}. 

We find that maximal compression is sub-optimal compared to the enhanced geometrical one. We speculate that the main reason behind this limitation is the linear limit implicit in the compression scheme we developed in \cite{Gualdi:2017iey,Gualdi:2019ybt}. With a non-linearly degenerate parameter space, even if these linear compression techniques achieve an improvement of the parameter constraints by allowing the employment of larger data-vectors, they still miss part of the available information. That could be the reason why combining the maximal compression with a complementary approach, such as the geometrical one, produces an effective non-linear compression which returns tighter parameters constraints once applied to an originally longer (but not redundant) data-vector.

This effective non-linearity feature is achieved by adopting a motivated criteria that defines which triangles are linearly combined together. The non-linearity indeed lies in the method, Eq. \ref{eq:final_bin_estimator}, used to define the triangle bins. In other words grouping the bispectra is non-linear operation in the cosmological parameters.

 With respect to the original data-vector, GEOMAX achieves a similar compression factor to the geometrical compression $\sim 23$, but a much lower one than maximal one ($\sim 683$).
 From the results displayed in Figure \ref{fig:data_mocks_4p}, we conclude that the reduction of the compressing factor is a fair price for the increased constraining power obtained by exploiting the physical insight granted by the geometrical compression step.

\section{Alternatives to the geometrical compression step}
\label{sec:alternatives_to_geo}
We show in Figure \ref{fig:alternatives_to_geo} two different ways to define the triangles sets, before applying the maximal compression in order to obtain the final data-vector. Since the full bispectrum for the standard MCMC $\Delta k_6$ case has 116 triangles, we group the 2734 triangles of the $\Delta k_2$ case into 29 sets. Maximal compression is then applied on each of these 29 groups. In this way, since we consider four parameters, the final data-vector will also have 116 elements. 

This is the largest allowed dimension in order to fairly compare the compressed data-vector (given by the $\Delta k_2$ triangles) with the original full one (given by the $\Delta k_6$ triangles).

\subsection{Random regrouping}
The most naive way to group together the triangles before the maximal compression step is to randomly distribute them into $N$-groups (as equally populated as possible). This immediately raises the concern that different random allocations of the triangles into $N$-groups can in principle produce wider/tighter posterior distributions. 

The performance cannot be predicted in advance unless it is applied an a-priori criteria to choose whether a random allocation is optimal or not. Even if such a criteria could be devised, the random choice factor would make its application very inefficient. On the contrary the geometrical step possesses a precise criteria to define the optimal way to group triangles together and also in how many bins. 

From the marginalised contours in Figure \ref{fig:alternatives_to_geo} we can immediately deduce that the geometrical compression algorithm outperforms this alternative. 

\subsection{Reference triangles}
A more sophisticated approach consists in defining "reference" triangles and to assign each of the $\Delta k_2$ case 2734 triangles to the bin defined by the most similar "reference" triangle. We then assign a triangle $a$ characterised by the sides $(k_1^a,k_2^a,k_3^a)$ to the bin $j$ having as reference triangle the configuration $(p_1^b,p_2^b,p_3^b)$ such that:

\begin{eqnarray}
\sum_{i=1,2,3} \dfrac{\abs{k^a_i - p^b_i}}{k^a_i} < 
\sum_{i=1,2,3} \dfrac{\abs{k^a_i - p^{\ell}_i}}{k^a_i}\;\forall\;\ell\neq b \,.
\end{eqnarray}

\noindent The similarity criteria is then simply the minimal sum of the normalised absolute difference between the triangles sides. The reference triangles are chosen among the original 2734 ones. Considering the algorithm that generated them, the selection is done by taking, in terms of the position in the array, equidistant configurations in the data-vector (one every 95 configurations in this particular case).

 Figure \ref{fig:alternatives_to_geo} shows that also this slightly more sophisticated criteria does not reach the improvements achieved by the enhanced geometrical compression. As in the case of the "random regrouping" approach, the marginalised posterior distributions are not significantly tighter than the one given by just using maximal compression.


\setlength{\tabcolsep}{4pt}
\begin{table}
	\centering
	  \caption[Improvement in parameter constraints]{Same as Table \ref{tab:improvements_4par} for the first additional parameter set test case.}
	\label{tab:improvements_5par}
  \begin{tabular}{ *{5 }{c} }
\toprule
& $\Delta\theta^{\mathrm{mc}}_{\Delta k_6}$ 
& \multicolumn{3}{c}{$\dfrac{ \Delta\theta^{\mathrm{mc}}_{\Delta k_6}-\Delta\theta^{\mathrm{comp.}}_{\Delta k_2}}{\Delta\theta^{\mathrm{mc}}_{\Delta k_6}}\;\left[\%\right]$}  \\
\cmidrule(lr){2-2}\cmidrule(lr){3-5}
        &  MCMC   & MAX  & GEO & GEOMAX \\
\cmidrule(lr){2-2}\cmidrule(lr){3-5}
        &  $N_{\mathrm{tr}}=116$   & $N_{\mathrm{el.}}=4$ &$N_{\mathrm{g}}=116$ &$N_{\mathrm{g}}=115$ \\
\cmidrule(lr){2-2}\cmidrule(lr){3-5}
$\Delta b_1 $          & 0.22 & \colorbox{-red!75!green}{36.1} &  \colorbox{yellow!70}{28.6} & \colorbox{red!75}{50.4}\\
$\Delta b_2 $          & 0.40 & \colorbox{-red!75!green}{45.5} &  \colorbox{yellow!70}{36.5} & \colorbox{red!75}{56.1}\\
$\Delta f   $          & 0.08 & \colorbox{-red!75!green}{23.9} &  \colorbox{yellow!70}{18.5} & \colorbox{red!75}{33.2}\\
$\Delta \sigma_8$      & 0.04 & \colorbox{-red!75!green}{21.6} &  \colorbox{yellow!70}{15.1} & \colorbox{red!75}{38.3}\\
$\Delta f_\mathrm{NL}$ & 171.5 & \colorbox{-red!75!green}{-5.8} &  \colorbox{yellow!70}{-6.1} & \colorbox{red!75}{6.8}\\
\cmidrule(lr){1-5}
\multicolumn{2}{c}{$\Big\langle\dfrac{
\Delta\theta^{\mathrm{mc}}_{\Delta k_6}-\Delta\theta^{\mathrm{comp.}}_{\Delta k_2}}{\Delta\theta^{\mathrm{mc}}_{\Delta k_6}}\;\left[\%\right]\Big\rangle$}
& 21.1 & 19.9 & 38.8\\

\bottomrule
    \end{tabular}
\end{table}


\setlength{\tabcolsep}{3pt}
\begin{table}
	\centering
	  \caption[Improvement in parameter constraints]{Same as Table \ref{tab:improvements_4par} for the second additional parameter set test case.}
	\label{tab:improvements_6par}
  \begin{tabular}{ *{5 }{c} }
& $\Delta\theta^{\mathrm{mc}}_{\Delta k_6}$ 
& \multicolumn{3}{c}{$\dfrac{ \Delta\theta^{\mathrm{mc}}_{\Delta k_6}-\Delta\theta^{\mathrm{comp.}}_{\Delta k_2}}{\Delta\theta^{\mathrm{mc}}_{\Delta k_6}}\;\left[\%\right]$}  \\
\cmidrule(lr){2-2}\cmidrule(lr){3-5}
        &  MCMC   & MAX  & GEO & GEOMAX \\
\cmidrule(lr){2-2}\cmidrule(lr){3-5}
        &  $N_{\mathrm{tr}}=116$   & $N_{\mathrm{el.}}=4$ &$N_{\mathrm{g}}=116$ &$N_{\mathrm{g}}=114$ \\
\cmidrule(lr){2-2}\cmidrule(lr){3-5}
$\Delta b_1 $                & 0.25 & \colorbox{-red!75!green}{30.2} &  \colorbox{yellow!70}{36.8} & \colorbox{red!75}{47.8}\\
$\Delta b_2 $                & 0.33 & \colorbox{-red!75!green}{36.0} &  \colorbox{yellow!70}{44.7} & \colorbox{red!75}{50.9}\\
$\Delta f   $                & 0.11 & \colorbox{-red!75!green}{27.0} &  \colorbox{yellow!70}{35.9} & \colorbox{red!75}{37.7}\\
$10^9\Delta A_{\mathrm{s}}$      & 0.27 & \colorbox{-red!75!green}{15.0} &  \colorbox{yellow!70}{4.7} & \colorbox{red!75}{32.0}\\
$\Delta \Omega_{\mathrm{m}}$ & 0.01 & \colorbox{-red!75!green}{-0.8} &  \colorbox{yellow!70}{3.6} & \colorbox{red!75}{8.8}\\
$\Delta f_\mathrm{NL}$ & 183 & \colorbox{-red!75!green}{-4.8} &  \colorbox{yellow!70}{-6.8} & \colorbox{red!75}{7.9}\\
\cmidrule(lr){1-5}
\multicolumn{2}{c}{$\Big\langle\dfrac{
\Delta\theta^{\mathrm{mc}}_{\Delta k_6}-\Delta\theta^{\mathrm{comp.}}_{\Delta k_2}}{\Delta\theta^{\mathrm{mc}}_{\Delta k_6}}\;\left[\%\right]\Big\rangle$}
& 17.1 & 19.8 & 30.8 \\
\bottomrule
    \end{tabular}
\end{table}

\section{Alternative parameter sets}
\label{sec:alternative_par_sets}
In order to check that the improvement achieved by the enhanced geometrical compression is not dependent on the chosen parameter set, we 
use GEOMAX to derive the constraints for two additional parameter sets from the same galaxy catalogues and data.

\subsection{Local primordial non-Gaussianity}
\label{sec:png}

In order to distinguish between different models of inflation, one of the key observables is the deviation from a Gaussian distribution of the primordial density fluctuations \citep{Bartolo:2004if}. In the case of local primordial non-Gaussianity this deviation can be parametrised through an expansion of the Bardeen's gravitational potential $\Phi$ \citep{Bardeen:1980kt} in terms of a Gaussian field $\phi$ and a costant $f_\mathrm{NL}$ acting as the amplitude of the deviation from linearity at first order:

\begin{eqnarray}
\Phi = \phi \,+\, \dfrac{f_{\mathrm{NL}}^2}{c^2}\left[\phi^2-\langle \phi^2\rangle\right]\,+ \,...
\end{eqnarray}

\noindent In our analysis we only consider a local type of primordial non-Gaussianity (see \citealp{Byrnes:2010em} for a review). There are also other types of primordial non-Gaussianities (for simulations on these see \citealp{Scoccimarro:2011pz}). 

Late time large-scale structure analyses have the potential of reaching, with the next generation of surveys, constraints on $f_{\mathrm{NL}}$ of similar order to the ones obtained by Planck \citep{Akrami:2019izv}. The modelling and constraints forecasts for the matter and galaxy bispectrum have already been widely studied in the literature \citep{Verde:1999ij,Sefusatti:2009qh,Jeong:2009vd}. We proceed as in \cite{Scoccimarro:2003wn} to derive the correction for the galaxy power spectrum for a local primordial non-Gaussianity:

\begin{align}
\label{eq:png_pk}
&\mathrm{P}^\mathrm{g}_{\mathrm{NG}}(\bm{k}) = \mathrm{P}_{11} + \mathrm{P}_{12} \approx \mathrm{P}_{12}
\notag\\
&=
\dfrac{4f_{\mathrm{NL}}}{c^2}F^{(1)}_k \beta^2 k^4\mathrm{T}_k^2\int\dfrac{d\bm{p}_a^3}{(2\pi)^3}F^{(2)}_{a|\bm{k}+\bm{p}_a|}
\mathrm{P}^{\phi}_{|\bm{k}+\bm{p}_a|}\left[\mathrm{P}^{\phi}_a+2\mathrm{P}^{\phi}_k\right] \,.
\end{align}

\noindent We only use $\mathrm{P}_{12}$ ($\propto {f_\mathrm{NL}}/{c^2}$) since $\mathrm{P}_{11}$ is expected to be negligible given that is proportional to ${f^2_\mathrm{NL}}/{c^4}$. In the above expression $F^{(1)}_k $ and $F^{(2)}_{a|\bm{k}+\bm{p}_a|}$ are the standard first and second order perturbation theory kernels in redshift space (know also as $Z$ in the literature). $\mathrm{T}_k$ is the matter transfer function normalised to one for $k\rightarrow 0$. $\mathrm{P}^{\phi}$ is the primordial power spectrum for the Gaussian part of the Bardeen's potential $\phi$. $\beta = \frac{3}{5}{D_1(z)}/\left({\Omega_\mathrm{m}H_0}\right)$ where $D_1$ is the linear growth factor at redshift $z$ while $\Omega_\mathrm{m}$ and $H_0$ are the matter density parameter and the Hubble constant, respectively. We fixed $H_0$ to the fiducial value used to compute the linear matter power spectrum.

For the bispectrum we have the additional term:
\begin{align}
\label{eq:png_bk}
&\mathrm{B}^\mathrm{g}_{\mathrm{NG}}(\bm{k}_1,\bm{k}_2,\bm{k}_3) = \mathrm{B}^\mathrm{g}_{111}
\notag \\
&=
F^{(1)}_{k_1}F^{(1)}_{k_2}F^{(1)}_{k_3}\beta^{-1}k_1^2k_2^{-2}k_3^{-2}\dfrac{\mathrm{T}_{k_1}}{\mathrm{T}_{k_2}\mathrm{T}_{k_3}}\dfrac{2f_{\mathrm{NL}}}{c^2}
\mathrm{P}^{\mathrm{m}}_{k_2}\mathrm{P}^{\mathrm{m}}_{k_3} \;+\; \mathrm{cyc.}\,,
\end{align}

\noindent where $\mathrm{P}^\mathrm{m}$ is the linear matter power spectrum. The derivation of both power spectrum and bispectrum primordial non-Gaussianity corrections is described in Appendix \ref{sec:appendix_png}. The relevance of the primordial non-Gaussianity terms with respect to the gravitational ones is shown in Figure \ref{fig:pk02_png} for the power spectrum and Figure \ref{fig:bk0_png} for the bispectrum.

Figure \ref{fig:data_mocks_5p} displays the results relative to the addition of the $f_{\mathrm{NL}}$ parameter to the analysis for both mocks and data measurements. 
While MAX and GEO compression return larger posterior distributions for $f_{\mathrm{NL}}$ than the standard MCMC on the full data-vector with less triangles, GEOMAX returns 1D $68\%$ credible regions tighter than the MCMC ones by $\sim10\%$ on average on mocks (left panel Figure \ref{fig:data_mocks_5p}) and $6.8\%$ for data (right panel Figure \ref{fig:data_mocks_5p} and Table \ref{tab:improvements_5par}), respectively.


\begin{figure}%
    \centering
    \includegraphics[width=0.45\textwidth]
    {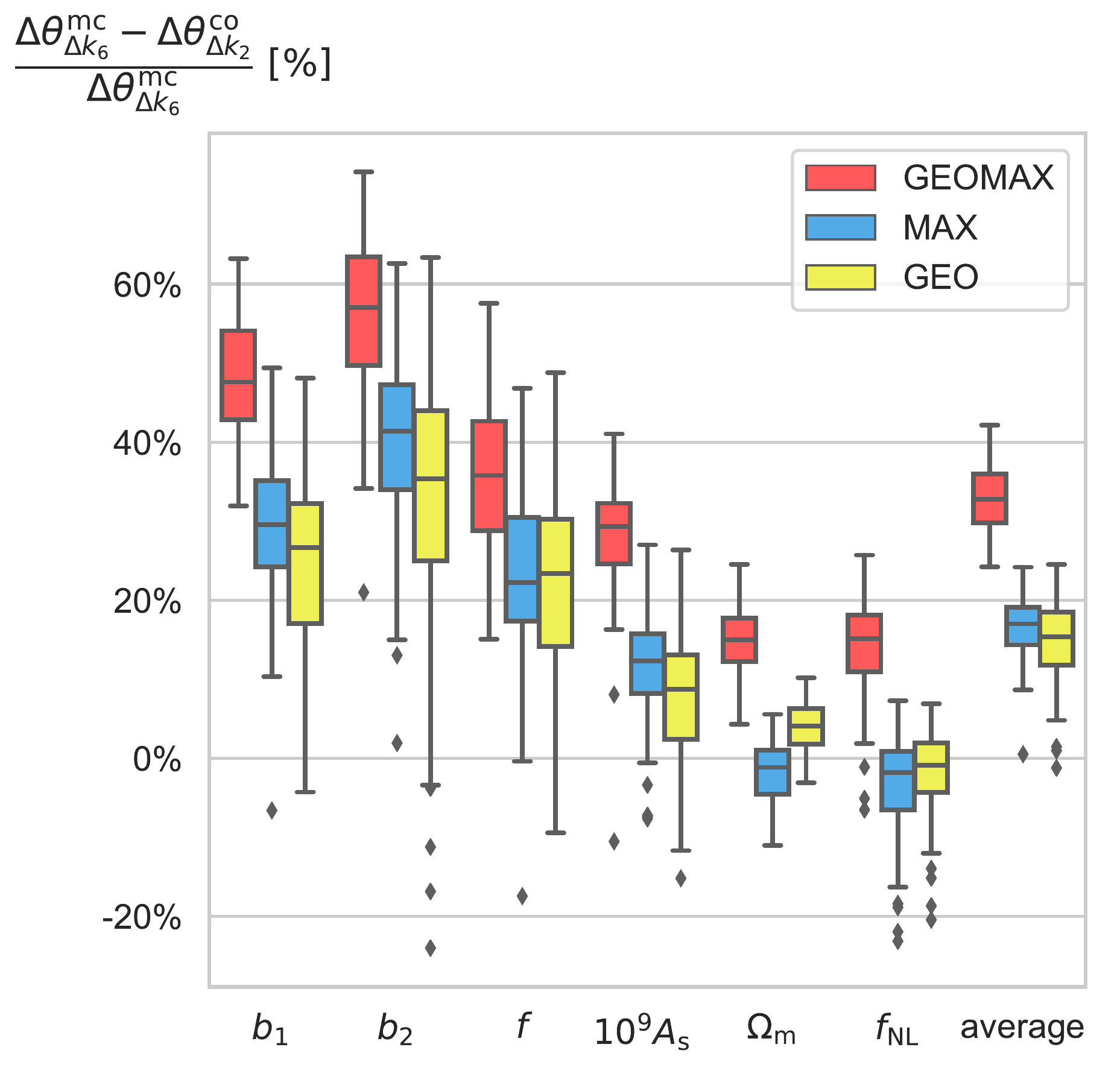}
    \caption{
    Same as right panel of Figure \ref{fig:data_mocks_4p} for the six parameters case. Also in this case the improvement of GEOMAX with respect to MAX and GEO compression methods is statistically significant.
}
    \label{fig:mocks_6p}
\end{figure}


\subsection{Cosmological parameters}
\label{sec:cosmo_par}
\begin{figure*}%
    \centering
    \includegraphics[width=0.9\textwidth]
    {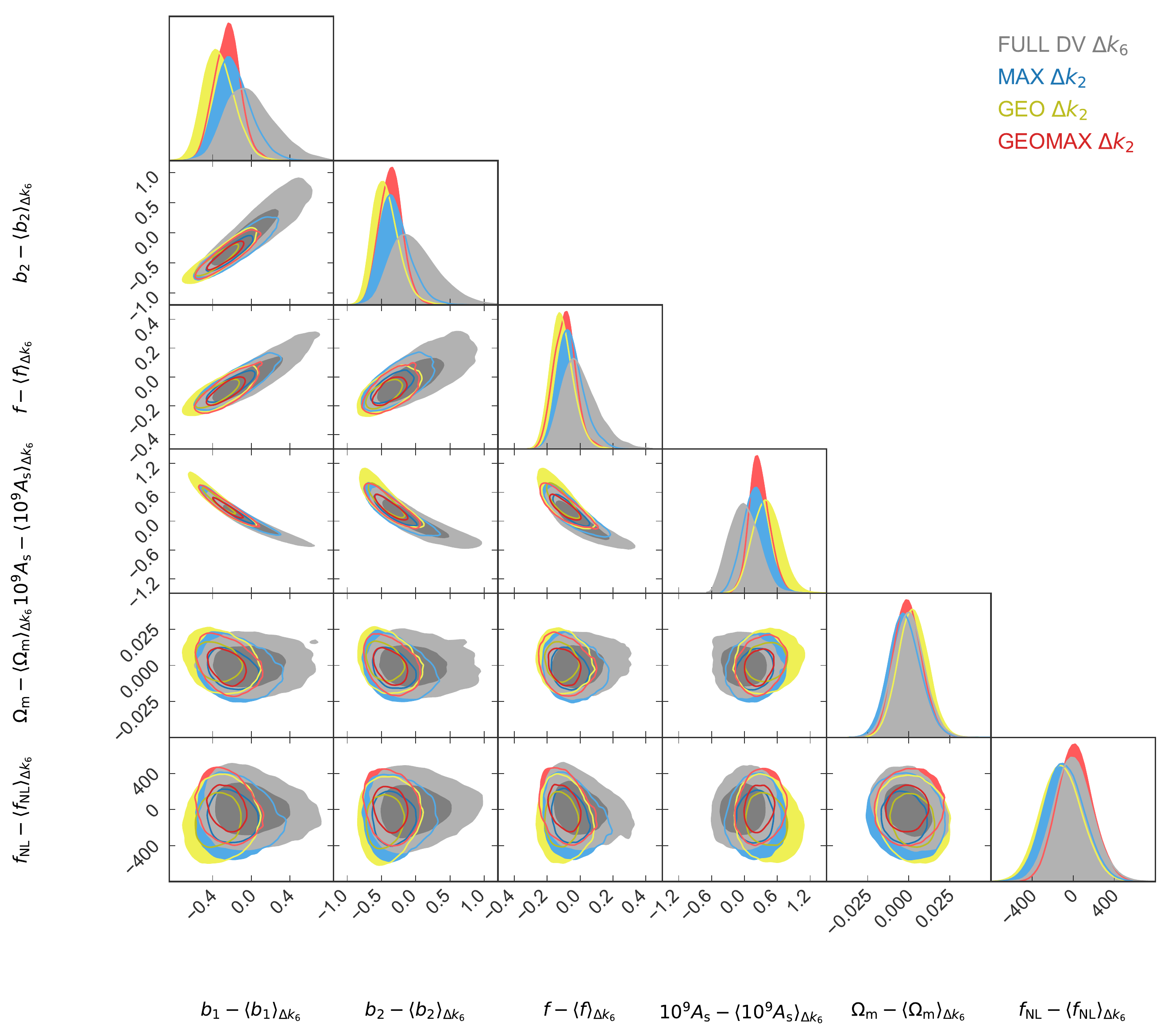}
    \caption{
    Same as left panel of Figure \ref{fig:data_mocks_4p} for the six parameters case. The enhanced geometrical compression still outperforms the individual maximal and geometrical ones. In particular it also improves the constraints for those parameters where the other methods performs as well as the MCMC on the full data-vector with less triangles.
}
    \label{fig:data_6p}
\end{figure*}

 In the second alternative parameter set we substitute $\sigma_8$ with the amplitude of scalar perturbations $A_{\mathrm{s}}$ and the matter density parameter $\Omega_{\mathrm{m}}$. The results for this case are displayed in Figures \ref{fig:mocks_6p} and \ref{fig:data_6p}.
 
 Once more GEOMAX outperforms both MAX and GEO methods over all the parameters, improving on average by $\sim30\%$ the constraints given by standard MCMC on the full data-vector (Table \ref{tab:improvements_6par}).
In particular this case again shows that the enhanced compression is able to obtain noticeable improvements for those parameters ( $\Omega_{\mathrm{m}}, f_\mathrm{NL}$ ) where maximal and geometrical compression do not surpass the results of standard MCMC sampling.


\section{Conclusions}
\label{sec:conclusions}

We introduced an optimised compression method for the galaxy bispectrum and we made publicly available the code\footnote{\url{https://github.com/davidegua/max_geo_compression.git}}. This enhanced geometrical compression (GEOMAX) can in principle be easily applied to any 3pt statistics in cosmology. 

The first requirement are a sufficient large number of simulations. These however are normally far fewer than the ones that would be needed to estimate the covariance matrix for a 3pt statistics data-vector without compression.
The second input are the derivatives of the data-vector model with respect to the model parameters that one wishes to constrain.

The geometrical compression (GEO, \citealp{Gualdi:2019ybt}) splits the triangles into groups based on similarity of their geometrical properties. This allows us to use the available simulations to estimate the covariance matrix for each of these groups, which can then be used to maximally compress (MAX, \citealp{Gualdi:2017iey,Gualdi:2018pyw}) the bispectrum in each of them.

We tested the GEOMAX algorithm on both a set of galaxy bispectrum monopole measurements from 100 Patchy mocks \citep{Kitaura:2015uqa} and the measurement from BOSS DR12 CMASS sample \citep{Gil-Marin:2016wya}. Through the galaxy catalogues we studied the statistical significance of the improvements on parameter constraints (left panel Figures \ref{fig:data_mocks_4p}, \ref{fig:data_mocks_5p} and Figure \ref{fig:mocks_6p}). 
In Figure \ref{fig:loss_info4p} we also show that GEOMAX is statistically consistent with being "information-lossless" with respect to the MCMC on the full data-vector (while GEO and MAX methods are not). It would be interesting to compare this method's performance with other information lossless compression algorithms \citep{Alsing:2017var,Charnock:2018ogm}, which are usually directly applied to data through likelihood-free inference analyses \citep{Alsing:2018eau,Alsing:2019xrx}.
 
We used the DR12 CMASS data measurements to check that known systematics (for example the choice of a fiducial cosmology for the analysis) and  unknown ones did not affect the compression results.
With respect to the standard MCMC on the full data-vector, the enhanced geometrical compression returns 1D 68$\%$ credible regions tighter by a factor of $\left(50.4\%,56.1\%,33.2\%,38.3\%\right)$ for the parameters $\left(b_1,b_2,f,\sigma_8\right)$. With respect to the individual maximal and geometrical compression methods the constraints are $\sim 15\%$ smaller (see Figure \ref{fig:data_mocks_4p} and Table \ref{tab:improvements_4par} for details). 

Two alternative pre-maximal compression steps have been considered in order to test the importance of the geometrical method. These alternatives do not produce the same improvements as the geometrical method when combined with the maximal compression (Figure \ref{fig:alternatives_to_geo}). Moreover they don't show significant differences from maximal compression alone.

To strengthen our case, we also run the analysis for two larger parameter sets, proving that the benefits of this new method are not parameter-set dependent (Figures \ref{fig:data_mocks_5p},\ref{fig:mocks_6p},\ref{fig:data_6p} and Tables \ref{tab:improvements_5par},\ref{tab:improvements_6par}). In particular GEOMAX improves the MCMC 1D 68$\%$ credible regions also for those parameters where instead MAX and GEO methods return larger marginalised 1D posterior distribution. 
For these two additional parameter sets, the average improvement observed on data of GEOMAX with respect to MAX and GEO methods varies between $10\%$ and $20\%$.

In order to maximise the extraction of cosmological information from 3pt statistics, we conclude with the expectation that this flexible method will be employed for the analysis of the forthcoming cosmological data-sets such as DESI, Euclid, PFS and SKA.

\section*{ACKNOWLEDGMENTS} 
D.G. thanks Prof. Alan Heavens, Prof. Andrew Pontzen and Prof. Licia Verde for the useful discussions. D.G. is also grateful to Pérez Forcadell Gabriel for the help in using the ICCC-UB computer cluster.
D.G. acknowledges support from European Union’s Horizon 2020 research and innovation programme ERC (BePreSySe, grant agreement 725327).
HGM acknowledges the support from la Caixa Foundation (ID 100010434) which code LCF/BQ/PI18/11630024.
 MM acknowledges support from the European Union’s Horizon 2020 research and innovation program under Marie Sklodowska-Curie grant agreement No 6655919y.

The linear matter power spectrum has been computed using the CLASS code \citep{Lesgourgues:2011re}.
C \citep{Kernighan:1988:CPL:576122} 
and \textsc{python} 2.7 \citep{Rossum:1995:PRM:869369} have been used together with many packages like I\textsc{python}s
\citep{Perez:2007:ISI:1251563.1251831}, Numpy \citep{DBLP:journals/corr/abs-1102-1523}, Scipy \citep{jones} and Matplotlib \citep{Hunter:2007:MGE:1251563.1251845}. The corner plots have been realised using 
\textsc{pygtc} developed by \cite{Bocquet2016}. We used Emcee \citep{2013PASP..125..306F} as MCMC sampler.
 

\section*{DATA AVAILABILITY}
The data underlying this article are available in \url{https://data.sdss.org/sas/dr12/boss/lss/} from the public domain source at \url{https://www.sdss.org/dr12/}. 



\bibliographystyle{mnras}

\bibliography{geo_max_lib}
 
\onecolumn

 \appendix

\section{Data vector models}
\label{sec:appendix_models}
In this section we list the analytical expressions we used for computing the various terms of the data vector. The monopole and quadrupole of the galaxy power spectrum are given by:

\begin{eqnarray}
\label{pk_02_esp}
 \mathrm{P}_{\mathrm{g}}^{(\ell)}\left(k\right) = \dfrac{2\ell+ 1}{2}\int^{+1}_{-1}d\mu\,\mathrm{P}_{\mathrm{g}}^{(s)}\left(k, \mu\right)L_{\ell}\left(\mu\right)\,,
\end{eqnarray}

\noindent where $L_{\ell}\left(\mu\right)$ is the $\ell$-order Legendre polynomial and $\mathrm{P}_{\mathrm{g}}^{(s)}\left(k, \mu\right)$ is the redshift space galaxy power spectrum at tree level \citep{Gualdi:2018pyw}.
For the bispectrum monopole we adopt the effective formula given by \cite{GilMarin:2011ik} which was calibrated on simulations.

\begin{eqnarray}
\label{bk_0_esp}
\mathrm{B}^{(0)}_{\mathrm{g}}\left(k_1,k_2,k_3\right)
&= \dfrac{1}{4}\int^{1}_{-1} d\mu_1\int^{1}_{-1}d\mu_2 \,\mathrm{B}^{(s)}_{\mathrm{g}}\left(\bm{k}_1,\bm{k}_2,\bm{k}_3\right)
\notag \\
&=\dfrac{1}{4\pi}\int^{1}_{-1} d\mu_1\int^{2\pi}_{0}d\phi \, \mathrm{B}^{(s)}_{\mathrm{g}}\left(\bm{k}_1,\bm{k}_2,\bm{k}_3\right)
\,,
\end{eqnarray}

\noindent where $\mu_i$ is the angle between the $\bm{k}_i$ vector and the line of sight. The angle $\phi$ is defined as $\mu_2\equiv\mu_1x_{12} - \sqrt{1-\mu_1^2}\sqrt{1-x_{12}^2}\cos{\phi}$, where $x_{12}$ is the cosine of the angle between $\bm{k}_1$ and $\bm{k}_2$.

\section{Primordial non-Gaussianity expansion}
\label{sec:appendix_png}
\noindent Following what done in \cite{Scoccimarro:2003wn} we can compute for the power spectrum and the bispectrum, the contribution due to the presence of a primordial non-Gaussian component in the potential field. In order to do so we assume a local type of non-Gaussianity which in terms of the primordial potential can be parametrised as:

\begin{eqnarray}
\Phi_{\mathrm{p}}(\bm{x})&=&\phi_{\mathrm{p}}(\bm{x})+\dfrac{f_{\mathrm{NL}}}{c^2}\left[\phi^2_{\mathrm{p}}(\bm{x})-\langle\phi^2_{\mathrm{p}}(\bm{x})\rangle\right]
+\dfrac{g_{\mathrm{NL}}}{c^4}\left[\phi^3_{\mathrm{p}}(\bm{x})-3\phi_{\mathrm{p}}(\bm{x})\langle\phi^2_{\mathrm{p}}(\bm{x})\rangle\right] \;+\;.\,.\,. 
\end{eqnarray}{}

\noindent where $\phi_{\mathrm{p}}$ represents a Gaussian field while $f_\mathrm{NL}$ and $g_\mathrm{NL}$ are the constant parameters of the expansion up to third order in $\phi_{\mathrm{p}}$. In Fourier space it translates to (dropping the "p" index for $\phi$):

\begin{eqnarray}
\Phi_{\mathrm{p}}(\bm{k})&=&\phi_{k}+\dfrac{f_{\mathrm{NL}}}{c^2}\left[ I^k_{ab}\phi_a\phi_b-\delta_D(\bm{k})\langle\phi^2\rangle\right]
+\dfrac{g_{\mathrm{NL}}}{c^4}\left[ I^k_{abc}\phi_a\phi_b\phi_c-\dfrac{3}{(2\pi)^3}\phi_k\langle\phi^2\rangle\right]\,,
\end{eqnarray}{}

\noindent where in Fourier space $\langle\phi^2\rangle=\int d\bm{q}^3\mathrm{P}_{\phi}(\bm{q}) = (2\pi)^3\sigma^2_{\phi}$ and $\delta_D(\bm{k})$ is the Dirac's delta function. We introduced the short notation for the integral over the wave-vectors:

\begin{eqnarray}
I^{k}_{ab} &=& \int\dfrac{d\bm{q}_a^3d\bm{q}_b^3}{(2\pi)^3}\,\delta_D(\bm{k}-\bm{q}_a-\bm{q}_b)\, \notag \\
I^{k}_{abc} &=& \int\dfrac{d\bm{q}_a^3d\bm{q}_b^3d\bm{q}_c^3}{(2\pi)^6}\,\delta_D(\bm{k}-\bm{q}_a-\bm{q}_b-\bm{q}_c)\,.
\end{eqnarray}{}

The primordial potential is related to the late-times one by:

\begin{eqnarray}
\Phi_{\mathrm{l.t.}}(a) = \dfrac{9}{10}\dfrac{D_+}{a}\mathrm{T}(k)\Phi_{\mathrm{p}}\,, 
\end{eqnarray}{}
\noindent where $D_+(a)$ is the growth factor from linear perturbation theory as a function of the scale factor $a$. T($k$) is the transfer function normalised to unity for $k\rightarrow0$. At late-times the potential field is related to the density perturbation variable by the Poisson equation:

\begin{eqnarray}
\nabla^2\Phi_{\mathrm{l.t.}}(\bm{x},a)=\dfrac{3}{2}\dfrac{\Omega_{\mathrm{m}}H_0^2}{a}\delta(\bm{x},a)\,. 
\end{eqnarray}

\noindent This allows to link the primordial potential with the late-times matter density perturbation:

\begin{eqnarray}
\delta_k = \dfrac{3}{5}\dfrac{D_+}{\Omega_{\mathrm{m}}H_0^2
}k^2\mathrm{T}_k\Phi_{\mathrm{p}}=\beta k^2\mathrm{T}_k\Phi_{\mathrm{p}}\,.
\end{eqnarray}{}

\subsection{Power spectrum}
Let's start with the two-point correlation function in Fourier space. We will consider all the terms up to order $\phi^4$.

\begin{eqnarray}
\langle\delta_\mathrm{s}\delta_\mathrm{s}\rangle = \langle(\delta_{\mathrm{m}}^{(1)}+\delta_{\mathrm{m}}^{(2)}+\delta_{\mathrm{m}}^{(3)}+O(\delta_{\mathrm{m}}^{(4)}))(\delta_{\mathrm{m}}^{(1)}+\delta_{\mathrm{m}}^{(2)}+\delta_{\mathrm{m}}^{(3)}+O(\delta_{\mathrm{m}}^{(4)}))\rangle\,,
\notag \\
\end{eqnarray}{}

\noindent where the upper index represents the order in terms of $\delta_{\mathrm{m}}$ given by the expansion done in the previous section. Up to the considered order we then have:

\begin{eqnarray}
\langle\delta_\mathrm{s}\delta_\mathrm{s}\rangle &=&
\langle\delta_{\mathrm{m}}^{(1)}\delta_{\mathrm{m}}^{(1)}\rangle + 
2\langle\delta_{\mathrm{m}}^{(1)}\delta_{\mathrm{m}}^{(2)}\rangle + 
\langle\delta_{\mathrm{m}}^{(2)}\delta_{\mathrm{m}}^{(2)}\rangle + 
2\langle\delta_{\mathrm{m}}^{(1)}\delta_{\mathrm{m}}^{(3)}\rangle \notag \\
&=&\mathrm{P}_{11}+\mathrm{P}_{12}+\mathrm{P}_{22}+\mathrm{P}_{13}\,.
\end{eqnarray}{}

\subsubsection{$\mathrm{P}_{11}$}
Expanding in term of the primordial Gaussian potential $\phi$:

\begin{eqnarray}
\langle\delta_{\mathrm{m}}^{(1)}(\bm{k})\delta_{\mathrm{m}}^{(1)}(\bm{q})\rangle &=&\langle F^{(1)}_kF^{(1)}_q\beta^2k^2q^2\mathrm{T}_k\mathrm{T}_q 
\notag \\
&\times&\left\{ \phi_{k}+\dfrac{f_{\mathrm{NL}}}{c^2}\left[ I^k_{ab}\phi_a\phi_b-\delta_D(\bm{k})\langle\phi^2\rangle\right]
+\dfrac{g_{\mathrm{NL}}}{c^4}\left[ I^k_{def}\phi_d\phi_e\phi_f-\dfrac{3}{(2\pi)^3}\phi_k\langle\phi^2\rangle\right] \right\}
\notag \\
&\times&\left\{\phi_{q}+\dfrac{f_{\mathrm{NL}}}{c^2}\left[ I^q_{gh}\phi_g\phi_h-\delta_D(\bm{q})\langle\phi^2\rangle\right]
+\dfrac{g_{\mathrm{NL}}}{c^4}\left[ I^q_{ilm}\phi_i\phi_l\phi_m-\dfrac{3}{(2\pi)^3}\phi_q\langle\phi^2\rangle\right]\right\}\rangle\,.
\end{eqnarray}{}

\noindent Recalling that all odd moments of a Gaussian variable ($\phi$) are equal to zero and ignoring all higher order terms ($>\phi^4$) we obtain:

\begin{eqnarray}
\label{eq:png_p11}
&&\langle\delta_{\mathrm{m}}^{(1)}(\bm{k})\delta_{\mathrm{m}}^{(1)}(\bm{q})\rangle
= \langle F^{(1)}_kF^{(1)}_q\beta^2k^2q^2\mathrm{T}_k\mathrm{T}_q \times
\notag \\
&&\Bigg\{\phi_k\phi_q 
+ \dfrac{f_{\mathrm{NL}}^2}{c^4}\left[ I^k_{ab}\phi_a\phi_b-\delta_D(\bm{k})\langle\phi^2\rangle\right]\times
\left[ I^q_{gh}\phi_g\phi_h-\delta_D(\bm{q})\langle\phi^2\rangle\right]
+ 2\phi_k\dfrac{g_{\mathrm{NL}}}{c^4}\left[ I^q_{ilm}\phi_i\phi_l\phi_m-\dfrac{3}{(2\pi)^3}\phi_q\langle\phi^2\rangle\right]
\Bigg\}\rangle
\notag \\[2mm]
&&=F^{(1)}_kF^{(1)}_q\beta^2k^2q^2\mathrm{T}_k\mathrm{T}_q\times
\Bigg\{(2\pi)^3\delta_D^{kq}\mathrm{P}^{\phi}_k+
\dfrac{f_{\mathrm{NL}}^2}{c^4}\left[I^k_{ab}I^q_{gh}\langle\phi_a\phi_b\phi_g\phi_h\rangle - 2I^k_{ab}\langle\phi_a\phi_b\rangle\delta_D^q\langle\phi^2\rangle+\delta_D^k\delta_D^q\langle\phi^2\rangle^2\right]
\notag \\
&&+\dfrac{g_{\mathrm{NL}}}{c^4}\left[ I^q_{ilm}\langle\phi_i\phi_l\phi_m\phi_k\rangle -\dfrac{3}{(2\pi)^3}\langle\phi_k\phi_q\rangle\langle\phi^2\rangle\right]
\Bigg\}
\notag \\[2mm]
&&=F^{(1)}_kF^{(1)}_q\beta^2k^2q^2\mathrm{T}_k\mathrm{T}_q
\notag \\
&&\times\Bigg\{(2\pi)^3\delta_D^{kq}\mathrm{P}^{\phi}_k
+\dfrac{f_{\mathrm{NL}}^2}{c^4}\left[2I^k_{ab}(2\pi)^3\delta_D^{abq}\mathrm{P}^{\phi}_a\mathrm{P}^{\phi}_b
+\delta^k_D\delta^q_D\langle\phi^2\rangle^2-2\delta^k_D\delta^q_D\langle\phi^2\rangle^2+\delta^k_D\delta^q_D\langle\phi^2\rangle^2\right]
\notag \\
&&+\dfrac{g_{\mathrm{NL}}}{c^4}\left[
I^q_{ilm}3\langle\phi_i\phi_l\rangle\langle\phi_m\phi_k\rangle -3\delta_D^{kq}\mathrm{P}^{\phi}_k\langle\phi^2\rangle\right]\Bigg\}
\notag \\
&&=F^{(1)}_kF^{(1)}_q\beta^2k^2q^2\mathrm{T}_k\mathrm{T}_q(2\pi)^3\delta_D^{kq}
\notag \\
&&\times
\Bigg\{\mathrm{P}^{\phi}_k
+\dfrac{2f_{\mathrm{NL}}^2}{c^4}\int \dfrac{d\bm{p}_a^3}{(2\pi)^3}\mathrm{P}^{\phi}_a\mathrm{P}^{\phi}_{|\bm{k}-\bm{p}_a|}
+\dfrac{g_{\mathrm{NL}}}{c^4(2\pi)^3}\left[3\mathrm{P}^{\phi}_k\int d\bm{p}_l^3\mathrm{P}_{l}^{\phi}-3\mathrm{P}^{\phi}_k\langle\phi^2\rangle\right]\Bigg\}
\notag \\
&&=
F^{(1)}_kF^{(1)}_q\beta^2k^2q^2\mathrm{T}_k\mathrm{T}_q(2\pi)^3\delta_D^{kq}
\Bigg\{\mathrm{P}^{\phi}_k
+\dfrac{2f_{\mathrm{NL}}^2}{c^4}\int \dfrac{d\bm{p}_a^3}{(2\pi)^3}\mathrm{P}^{\phi}_a\mathrm{P}^{\phi}_{|\bm{k}-\bm{p}_a|}\Bigg\}\,,
\end{eqnarray}{}

\noindent where $\delta_D^{kq} =\delta_D(\bm{k}+\bm{q})$.

\subsubsection{$\mathrm{P}_{12}$}
\begin{eqnarray}
\label{eq:png_p12}
2\langle\delta_{\mathrm{m}}^{(1)}(\bm{k})\delta_{\mathrm{m}}^{(2)}(\bm{q})\rangle
&=& \langle2 F^{(1)}_k I^q_{ab}F^{(2)}_{ab}\delta_k^{\ell}\delta_a^{\ell}\delta_b^{\ell}\rangle 
\notag \\
&=&\langle 2F^{(1)}_k \beta^3 k^2\mathrm{T}_k 
I^q_{ab}
q_a^2\mathrm{T}_aq_b^2\mathrm{T}_b
F^{(2)}_{ab}
\notag \\[2mm]
&\times&\left\{ \phi_{k}+\dfrac{f_{\mathrm{NL}}}{c^2}\left[ I^k_{cd}\phi_c\phi_d-\delta_D(\bm{k})\langle\phi^2\rangle\right]
+\dfrac{g_{\mathrm{NL}}}{c^4}\left[ I^k_{def}\phi_e\phi_f\phi_g-\dfrac{3}{(2\pi)^3}\phi_k\langle\phi^2\rangle\right] \right\}
\notag \\[2mm]
&\times&\left\{ \phi_{a}+\dfrac{f_{\mathrm{NL}}}{c^2}\left[ I^a_{hi}\phi_h\phi_i-\delta_D(\bm{p}_a)\langle\phi^2\rangle\right]
+\dfrac{g_{\mathrm{NL}}}{c^4}\left[ I^a_{lmn}\phi_l\phi_m\phi_n-\dfrac{3}{(2\pi)^3}\phi_a\langle\phi^2\rangle\right] \right\}
\notag \\
&\times&\left\{\phi_{b}+\dfrac{f_{\mathrm{NL}}}{c^2}\left[ I^b_{or}\phi_o\phi_r-\delta_D(\bm{p}_b)\langle\phi^2\rangle\right]
+\dfrac{g_{\mathrm{NL}}}{c^4}\left[ I^b_{stv}\phi_s\phi_t\phi_v-\dfrac{3}{(2\pi)^3}\phi_b\langle\phi^2\rangle\right]\right\}\rangle\,.
\end{eqnarray}{}

\noindent This at maximum order $\phi^4$ returns only one term proportional to $f_{NL}$:

\begin{eqnarray}
&&2\langle\delta_{\mathrm{m}}^{(1)}(\bm{k})\delta_{\mathrm{m}}^{(2)}(\bm{q})\rangle
=2F^{(1)}_k \beta^3 k^2\mathrm{T}_k 
 I^q_{ab}
 q_a^2\mathrm{T}_aq_b^2\mathrm{T}_b
 F^{(2)}_{ab}
\dfrac{f_{\mathrm{NL}}}{c^2}
\notag \\
&&\times
\left\{I^k_{cd}\langle\phi_c\phi_d\phi_a\phi_b\rangle - \langle\phi_c\phi_d\rangle\delta_D(\bm{k})\langle\phi^2\rangle + 2I^a_{hi}\langle\phi_k\phi_b\phi_h\phi_i\rangle-2\langle\phi_h\phi_i\rangle\delta_D(\bm{p}_a)\langle\phi^2\rangle 
\right\}
\notag \\
&&=(2\pi)^6
\dfrac{4f_{\mathrm{NL}}}{c^2}F^{(1)}_k \beta^3 k^2\mathrm{T}_k
I_{ab}^qq_a^2\mathrm{T}_aq_b^2\mathrm{T}_bF^{(2)}_{ab}
\times
\left\{I_{cd}^k
\delta_D^{ac}\delta_D^{bd}\mathrm{P}^{\phi}_a\mathrm{P}^{\phi}_b + 2I^a_{hi}
\delta_D^{ki}\delta_D^{bh}\mathrm{P}^{\phi}_k\mathrm{P}^{\phi}_b
\right\}
\notag \\
&&=(2\pi)^3\delta_D^{kq}\dfrac{4f_{\mathrm{NL}}}{c^2}F^{(1)}_k \beta^3 k^2\mathrm{T}_k
\int\dfrac{d\bm{p}_a^3}{(2\pi)^3}
p_a^2\mathrm{T}_a|-\bm{k}-\bm{p}_a|^2\mathrm{T}_{|-\bm{k}-\bm{p}_a|}
F^{(2)}_{a,-\bm{k}-\bm{p}_a}\mathrm{P}^{\phi}_{|-\bm{k}-\bm{p}_a|}\left[\mathrm{P}^{\phi}_a+2\mathrm{P}^{\phi}_k\right]\,.
\notag \\
\end{eqnarray}

\begin{figure}%
    \centering
    \includegraphics[width=0.95\textwidth]
    {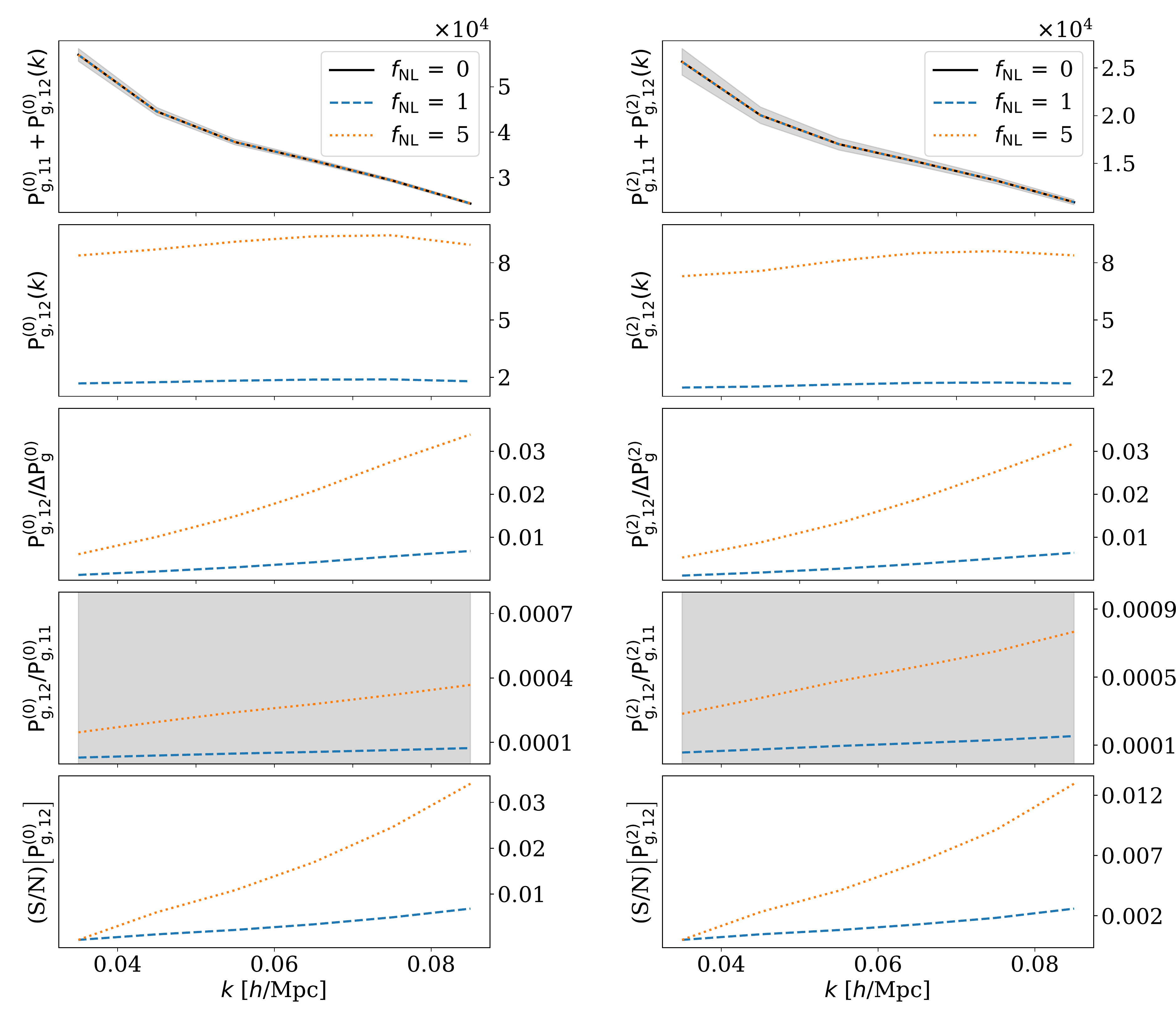}
    \caption{
    $\mathrm{P}_{\mathrm{g}}^{(0,2)}$: primordial non-Gaussianity contribution.
    In the first row the power spectrum monopole and quadrupole model is shown for different values of local primordial non-Gaussianity, $f_{\mathrm{NL}}=0,1,5$. The shaded area corresponds to the error-bars derived from the 1400 galaxy mocks measurements \citep{Kitaura:2015uqa}. The second row shows only the power spectrum terms linearly proportional to $f_\mathrm{NL}$. The third row shows the ratio between the primordial term and the data-point errorbar width, defined as $\Delta \mathrm{P}^{(0,2)}_\mathrm{g}(k_i)=\sqrt{\mathrm{Cov}_{ii}}$. The fourth row shows the ratio between the primordial term and the gravitational collapse one for the power spectrum. In the last row we show the primordial part comulative signal to noise as a function of the number of $k$-bins included in the analysis. This is defined as $(\mathrm{S}/\mathrm{N})\left[\mathrm{P}^{(0,2)}_{\mathrm{g},12}\right]_i = 
    \sqrt{\mathrm{P}^{(0,2),\intercal}_{\mathrm{g},12,i} \cdot \mathrm{Cov}^{-1}\cdot
    \mathrm{P}^{(0,2)}_{\mathrm{g},12,i}}$, where $\mathrm{P}^{(0,2)}_{\mathrm{g},12,i}$ is power spectrum monopole/quadrupole data-vector up to the $k$-bin $k_i$. $ \mathrm{Cov}^{-1}$ is the covariance matrix for the reduced data-vector  $\mathrm{P}^{(0,2)}_{\mathrm{g},12,i}$.
}
    \label{fig:pk02_png}
\end{figure}

\noindent The other terms in the power spectrum expansion, $\mathrm{P}_{22}$ and $\mathrm{P}_{13}$, are at first order already proportional to $\phi^4$ and therefore in our case they just return the standard loop correction terms for the Gaussian initial conditions. In Figure \ref{fig:pk02_png} the overall and relative effect of primordial non-Gaussianity can be observed for $f_{\mathrm{NL}=0,1,5}$ in comparison to the current error-bars. 


\subsection{Bispectrum}
Also for the bispectrum we limit the expansion to the terms proportional to $\phi^4$ so that we do not need to use fourth order perturbation theory. However for clarity we will list all terms up to order $\phi^6$ even if we are not going to compute them explicitly.

\begin{eqnarray}
\langle\delta_\mathrm{s}\delta_\mathrm{s}\delta_\mathrm{s}\rangle &=& \langle(\delta_{\mathrm{m}}^{(1)}+\delta_{\mathrm{m}}^{(2)}+\delta_{\mathrm{m}}^{(3)}+O(\delta_{\mathrm{m}}^{(4)}))(\delta_{\mathrm{m}}^{(1)}+\delta_{\mathrm{m}}^{(2)}+\delta_{\mathrm{m}}^{(3)}+O(\delta_{\mathrm{m}}^{(4)}))(\delta_{\mathrm{m}}^{(1)}+\delta_{\mathrm{m}}^{(2)}+\delta_{\mathrm{m}}^{(3)}+O(\delta_{\mathrm{m}}^{(4)}))\rangle 
\notag \\
&=& 
\langle\delta_{\mathrm{m}}^{(1)}\delta_{\mathrm{m}}^{(1)}\delta_{\mathrm{m}}^{(1)}\rangle
+ 
3\langle\delta_{\mathrm{m}}^{(1)}\delta_{\mathrm{m}}^{(1)}\delta_{\mathrm{m}}^{(2)}\rangle 
\notag \\
&+&
3\langle\delta_{\mathrm{m}}^{(2)}\delta_{\mathrm{m}}^{(2)}\delta_{\mathrm{m}}^{(1)}\rangle 
+ 
3\langle\delta_{\mathrm{m}}^{(1)}\delta_{\mathrm{m}}^{(1)}\delta_{\mathrm{m}}^{(3)}\rangle
\notag \\
&+&
3\langle\delta_{\mathrm{m}}^{(1)}\delta_{\mathrm{m}}^{(1)}\delta_{\mathrm{m}}^{(4)}\rangle + 6\langle\delta_{\mathrm{m}}^{(1)}\delta_{\mathrm{m}}^{(2)}\delta_{\mathrm{m}}^{(3)}\rangle+\langle\delta_{\mathrm{m}}^{(2)}\delta_{\mathrm{m}}^{(2)}\delta_{\mathrm{m}}^{(2)}\rangle
\notag \\
&=&
\mathrm{B}_{111}+\mathrm{B}_{112}+\mathrm{B}_{122}+\mathrm{B}_{113}+\mathrm{B}_{114}+\mathrm{B}_{123}+\mathrm{B}_{222}\,. 
\end{eqnarray}{}

\noindent From the expansion above we can see that if we would only consider terms up to order $\phi^6$, then $\mathrm{B}_{114}$, $\mathrm{B}_{123}$ and $\mathrm{B}_{222}$ would just result into loop-corrections without primordial non-Gaussianity contributions. From $\mathrm{B}_{122}$ and $\mathrm{B}_{113}$ we would have only terms proportional to $f_{\mathrm{NL}}$ since the ones given by Gaussian initial conditions are all equal to zero (odd-moments).

For $\mathrm{B}_{111}$ we would need to consider an additional parameter for the primordial non-Gaussianity contribution, in particular one proportional to $\phi^4$. The other terms up to order $\phi^6$ originating from $\mathrm{B}_{111}$ would be either proportional to $f_{\mathrm{NL}}^3$ or to the product $f_{\mathrm{NL}}g_{\mathrm{NL}}$. Limiting ourselves only at order $\phi^4$, from $\mathrm{B}_{111}$ we will obtain only one term proportional to $f_{\mathrm{NL}}$.

Finally $\mathrm{B}_{112}$ would return in the case of Gaussian initial conditions the standard tree level expression for the bispectrum. When PNG are considered up to order $\phi^6$, $\mathrm{B}_{112}$ contains also terms proportional to both $f_{\mathrm{NL}}^2$ and $g_{\mathrm{NL}}$.


\begin{figure}%
    \centering
    \includegraphics[width=0.90\textwidth]
    {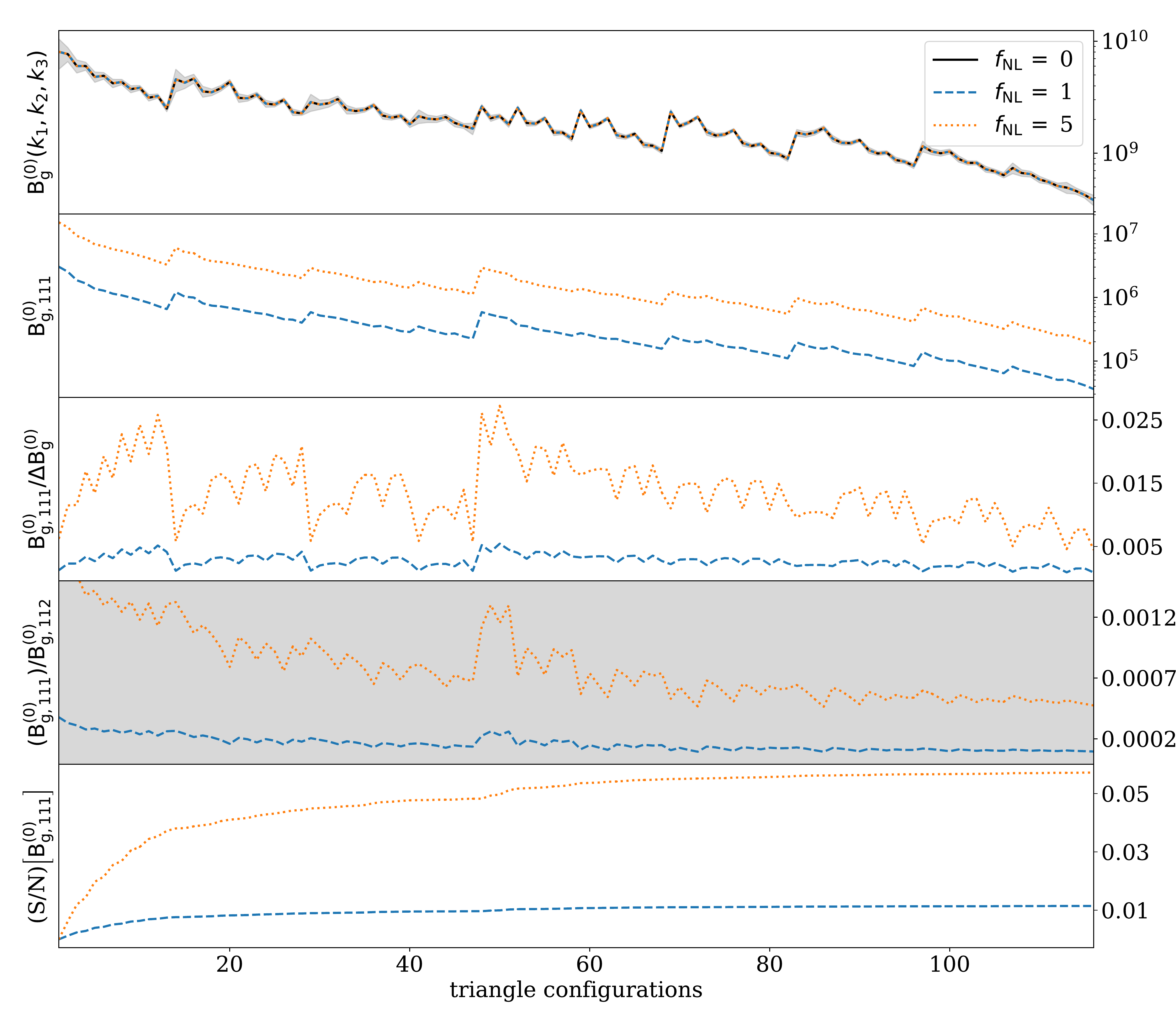}
    \caption{
$\mathrm{B}_{\mathrm{g}}^{(0)}$: primordial non-Gaussianity contribution. Same as Figure \ref{fig:pk02_png} for the galaxy bispectrum monopole.}
    \label{fig:bk0_png}
\end{figure}

\subsubsection{$\mathrm{B}_{111}$}
We proceed then with the only term containing primordial non-Gaussianity contributions up to order $\phi^4$:

\begin{eqnarray}
\langle\delta_{\mathrm{m}}^{(1)}\delta_{\mathrm{m}}^{(1)}\delta_{\mathrm{m}}^{(1)}\rangle &=& \langle F^{(1)}_{k_1}F^{(1)}_{k_2}F^{(1)}_{k_3}\beta^3k_1^2k_2^2k_3^2\mathrm{T}_{k_1}\mathrm{T}_{k_2}\mathrm{T}_{k_3}
\notag \\[2mm]
&\times&\left\{ \phi_{k_1}+\dfrac{f_{\mathrm{NL}}}{c^2}\left[ I^{k_1}_{ab}\phi_a\phi_b-\delta_D(\bm{k}_1)\langle\phi^2\rangle\right]
+\dfrac{g_{\mathrm{NL}}}{c^4}\left[ I^{k_1}_{cde}\phi_c\phi_d\phi_e-\dfrac{3}{(2\pi)^3}\phi_{k_1}\langle\phi^2\rangle\right] \right\}
\notag \\[2mm]
&\times&\left\{ \phi_{k_2}+\dfrac{f_{\mathrm{NL}}}{c^2}\left[ I^{k_2}_{fg}\phi_f\phi_g-\delta_D(\bm{k}_2)\langle\phi^2\rangle\right]
+\dfrac{g_{\mathrm{NL}}}{c^4}\left[ I^{k_2}_{hil}\phi_h\phi_i\phi_l-\dfrac{3}{(2\pi)^3}\phi_{k_2}\langle\phi^2\rangle\right] \right\}
\notag \\
&\times&\left\{\phi_{k_3}+\dfrac{f_{\mathrm{NL}}}{c^2}\left[ I^{k_3}_{mn}\phi_m\phi_n-\delta_D(\bm{k}_3)\langle\phi^2\rangle\right]
+\dfrac{g_{\mathrm{NL}}}{c^4}\left[ I^{k_3}_{oqr}\phi_o\phi_q\phi_r-\dfrac{3}{(2\pi)^3}\phi_{k_3}\langle\phi^2\rangle\right]\right\}\rangle\,.
\end{eqnarray}{}

\noindent The resulting three terms proportional to $f_{NL}$ are equivalent, different only by the permutation between $k_1$, $k_2$ and $k_3$:

\begin{eqnarray}
\langle\delta_{\mathrm{m}}^{(1)}\delta_{\mathrm{m}}^{(1)}\delta_{\mathrm{m}}^{(1)}\rangle &=& F^{(1)}_{k_1}F^{(1)}_{k_2}F^{(1)}_{k_3}\beta^3k_1^2k_2^2k_3^2\mathrm{T}_{k_1}\mathrm{T}_{k_2}\mathrm{T}_{k_3}\dfrac{f_{\mathrm{NL}}}{c^2}
\notag \\[2mm]
&\times& \left\{ I^{k_1}_{ab}\langle\phi_{k_2}\phi_{k_3}\phi_{a}\phi_{b}\rangle - \delta_D(\bm{k}_1)\langle\phi_{k_2}\phi_{k_3}\rangle\langle\phi^2\rangle\right\}
\;+\; \mathrm{cyc.}
\notag \\
&=&
F^{(1)}_{k_1}F^{(1)}_{k_2}F^{(1)}_{k_3}\beta^3k_1^2k_2^2k_3^2\mathrm{T}_{k_1}\mathrm{T}_{k_2}\mathrm{T}_{k_3}\dfrac{f_{\mathrm{NL}}}{c^2}
I^{k_1}_{ab}2(2\pi)^6\delta_D^{k_2a}\delta_D^{k_3b}\mathrm{P}^{\phi}_{k_2}\mathrm{P}^{\phi}_{k_3} \;+\; \mathrm{cyc.} 
\notag \\
&=&
(2\pi)^3F^{(1)}_{k_1}F^{(1)}_{k_2}F^{(1)}_{k_3}\beta^3k_1^2k_2^2k_3^2\mathrm{T}_{k_1}\mathrm{T}_{k_2}\mathrm{T}_{k_3}\dfrac{2f_{\mathrm{NL}}}{c^2}
\delta_D(\bm{k}_1+\bm{k}_2+\bm{k}_3)\mathrm{P}^{\phi}_{k_2}\mathrm{P}^{\phi}_{k_3} \;+\; \mathrm{cyc.}
\notag\\
&=&
(2\pi)^3\delta_D(\bm{k}_1+\bm{k}_2+\bm{k}_3)
F^{(1)}_{k_1}F^{(1)}_{k_2}F^{(1)}_{k_3}\beta^{-1}k_1^2k_2^{-2}k_3^{-2}\dfrac{\mathrm{T}_{k_1}}{\mathrm{T}_{k_2}\mathrm{T}_{k_3}}\dfrac{2f_{\mathrm{NL}}}{c^2}
P^{\mathrm{m}}_{k_2}P^{\mathrm{m}}_{k_3} \;+\; \mathrm{cyc.}\,,
\notag \\
\end{eqnarray}

\noindent where in the last line the primordial power spectrum was converted into the late-time matter power spectrum. The primordial non-Gaussianity contribution to the galaxy bispectrum monopole is shown in Figure \ref{fig:bk0_png}.

\section{Outliers analysis}
\label{sec:outliers}
In the left panels of Figures \ref{fig:data_mocks_4p}, \ref{fig:data_mocks_5p} and Figure \ref{fig:mocks_6p} we can observe that for few galaxy catalogues the relative parameter constraints improvements are negative. In the same figures we also check that the overall average improvement on the parameters constraints is always positive. 

\begin{figure}%
    \centering
    \includegraphics[width=0.90\textwidth]
    {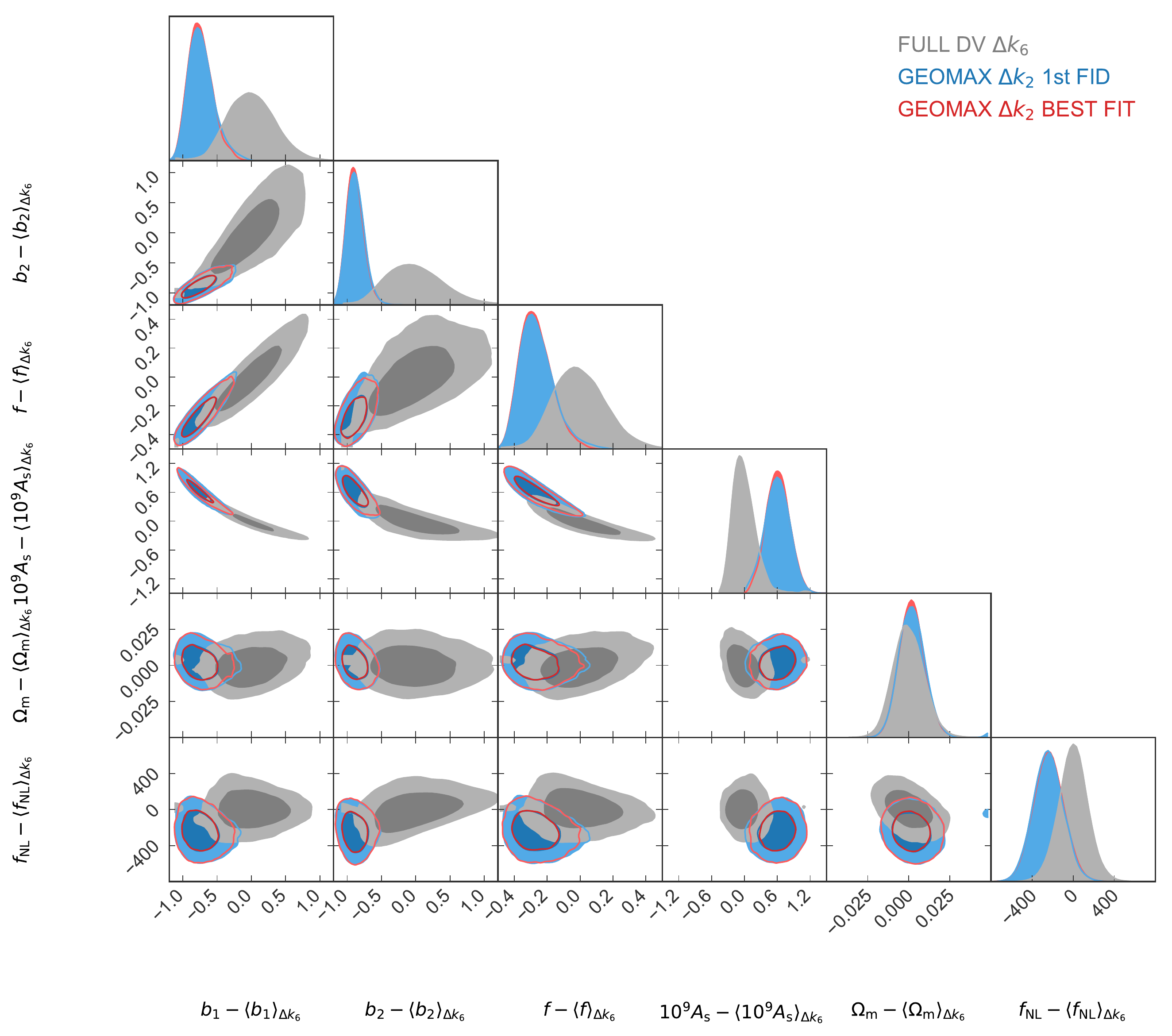}
    \caption{Posterior for the galaxy catalogue 37: beside the contours relative to the MCMC on the full data-vector containing 116 triangle configurations we show the result of GEOMAX compression applied to 2734 triangles bispectra. In the first case ("FID") we used the fiducial cosmology to compute the derivatives needed for the compression. The second set of contours ("BEST FIT") where instead derived by  using the best fit parameters obtained through the first run ("FID") to compute the derivatives needed in the geometrical step. Even if these two sets have differences larger than $1\sigma$ intervals for certain parameters, the 1-2D posterior contours given by GEOMAX  do not significantly differ.}
    \label{fig:mock37}
\end{figure}

We show in Figure \ref{fig:mock37} the 1-2D marginalised posterior distribution for one of the galaxy catalogues showing this negative improvement for some of the constrained parameters. In particular we show the posterior for the MCMC on the full data-vector (including 116 triangle configurations) and two different runs of the GEOMAX algorithm applied to the larger set of triangles (2734).
The first run is relative to the compression being computed using the same fiducial cosmology used throghout the paper. 
For the second run, we computed the compression using the best fit parameters set obtained from the first run.

Indeed we wanted to check whether the observed negative improvement for certain parameters was due to the difference between the assumed fiducial cosmology and the best fit one. This is not the case.
Nonetheless in this way we verified on this single case that the GEOMAX compression performance is not sensitive to the fiducial cosmology used to derive the compression weights.

We then conclude that the negative improvement observed for certain parameters is a statistical effect counterbalancing the above average improvement for the rest of the parameters. This second hypothesis is supported by the average column in the left panels of Figures \ref{fig:data_mocks_4p}, \ref{fig:data_mocks_5p} and Figure \ref{fig:mocks_6p}.


\bsp	
\label{lastpage}

\end{document}